\documentclass[a4paper,11pt]{article}
\pdfoutput=1 
\usepackage{jcappub} %
\usepackage{microtype}
\usepackage[T1]{fontenc} 
\usepackage{
amsmath, 
amssymb, 
amsthm, 
graphicx, 
epsfig, 
fancyhdr,
epsfig, 
slashed, 
mathrsfs,
mathtools,
physics
}
\usepackage{tikzsymbols}
\usepackage{natbib}
\usepackage{float}
\usepackage[thinc]{esdiff}
\usepackage{caption}
\usepackage{subcaption}
\usepackage{enumitem}
\usepackage{xcolor}
\usepackage{aas_macros}
\usepackage{acronym}
\usepackage{comment}
\usepackage[normalem]{ulem}

\usepackage{tikz,hyperref}
\hypersetup{
    colorlinks=true,
    citecolor=blue,
    linkcolor=brown,
    filecolor=magenta,      
    urlcolor=blue,
    pdftitle={latticePFLRW},
    pdfpagemode=UseOutlines,
}
\definecolor{lime}{HTML}{A6CE39}
\DeclareRobustCommand{\orcidicon}{
	\begin{tikzpicture}
	\draw[lime, fill=lime] (0,0) 
	circle [radius=0.2] 
	node[white] {{\fontfamily{qag}\selectfont \tiny ID}};
	\draw[white, fill=white] (-0.0625,0.095) 
	circle [radius=0.007];
	\end{tikzpicture}
	\hspace{-2mm}
}

\foreach \x in {A, ..., Z}{\expandafter\xdef\csname orcid\x\endcsname{\noexpand\href{https://orcid.org/\csname orcidauthor\x\endcsname}
			{\noexpand\orcidicon}}
}

\newcommand{\be}{\begin{equation}}
\newcommand{\ee}{\end{equation}}
\newcommand{\bea}{\begin{eqnarray}}
\newcommand{\eea}{\end{eqnarray}}

\def\Mpl{M_{\rm Pl}}

\newcommand{\ti}{t_{\rm i}}
\newcommand{\tf}{t_{\rm f}}

\newcommand{\stkout}[1]{\ifmmode\text{\sout{\ensuremath{#1}}}\else\sout{#1}\fi}

\def\Ng{N_{\!g}}

\newcommand{\eg}{e.g.\xspace}

\newcommand{\efold}{$e$-fold\xspace}
\newcommand{\efolds}{$e$-folds\xspace}
\newcommand{\efolding}{$e$-folding\xspace}

\newcommand{\est}{\mathrm{est}}
\newcommand{\fric}{\mathrm{fric}}

\newcommand{\calR}{\mathcal{R}}

\def\hmath$#1${\texorpdfstring{{\rmfamily\textit{#1}}}{#1}}
\definecolor{cardinal}{rgb}{0.77, 0.12, 0.23}

\title{
\boldmath
Nonlinear Lattice Framework for Inflation: Bridging stochastic inflation and the $\delta{N}$ formalism
}

\author[a]{Pankaj~Saha\orcidA{}}
\author[b]{Yuichiro~Tada\orcidB{}}
\author[a,c]{Yuko~Urakawa\orcidC{}}

\affiliation[a]{Theory Center, Institute of Particle and Nuclear Studies~(IPNS),\\~High Energy Accelerator
Research Organization (KEK),\\~Oho 1-1, Tsukuba 305-0801, Japan}
\affiliation[b]{Department of Physics, Rikkyo University, \\ 
3-34-1 Nishi-Ikebukuro, Toshima, Tokyo 171-8501, Japan}
\affiliation[c]{The Graduate University for Advanced Studies (SOKENDAI),\\~Tsukuba 305-0801, Japan}
\emailAdd{pankaj@post.kek.jp}
\emailAdd{yuichiro.tada@rikkyo.ac.jp}
\emailAdd{yukour@post.kek.jp}

\abstract{
Understanding when inflationary perturbations become genuinely nonlinear near the horizon crossing requires methods that go beyond both linear perturbation theory and the gradient expansion.
In this work, we introduce a nonlinear lattice framework for single-field inflation
based on a shear-free, locally \acl{FLRW} geometry.
This approach captures the leading effects of inhomogeneous local expansion rates,
curvature contributions to the local Friedmann equation, and proper-volume weighting at a fraction of the computational cost of full numerical relativity. 
We construct fully nonlinear $\delta N$ observables on uniform-density slices, together with other practical time-dependent estimators for the curvature perturbations.
After validating the framework in a standard slow-roll regime, we apply it to 
Starobinsky's linear-potential model featuring an intermittent \ac{USR} phase and a sharp potential transition. 
During this non-attractor USR regime, the lattice captures the separation of curvature perturbation estimators, the growth and subsequent stabilisation of non-Gaussianity, and a transient weakening of the shear-free approximation when the inflaton velocity becomes very small.
Our framework provides a practical intermediate approach between rigid background lattice simulations and full numerical relativity, offering a nonlinear bridge between lattice methods, the $\delta N$ formalism, and the stochastic inflation formalism as perturbations transition beyond the linear regime.
}

\acrodef{USR}{ultra-slow-roll}
\acrodef{BSSN}{Baumgarte--Shapiro--Shibata--Nakamura}
\acrodef{GR}{general relativity}
\acrodef{FLRW}{Friedmann--Lema\^itre--Robertson--Walker}
\acrodef{PBH}{primordial black hole}
\acrodef{CMB}{cosmic microwave background}
\acrodef{STOLAS}{STOchastic LAttice Simulation}
\acrodef{PDF}{probability density function}
\acrodef{rms}{root-mean-square}
\acrodef{FFT}{fast Fourier transform}
\acrodef{MS}{Mukhanov--Sasaki}
\acrodef{KG}{Klein--Gordon}

\begin{document} 
\noindent KEK-TH-2821\\
\noindent KEK-Cosmo-0415\\
\noindent RUP-26-4
\maketitle
\flushbottom
\acresetall

\section{\label{sec:intro}Introduction}

The observed temperature anisotropies in the \ac{CMB}, at the level of $\Delta T/T \sim 10^{-5}$~\cite{WMAP:2003ivt,WMAP:2003syu,Planck:2013pxb,BICEP2:2015nss,Planck:2018vyg}, imply that the Universe was remarkably homogeneous and isotropic at the time of recombination across all scales probed by the CMB{~---~ranging from $\mathcal{O}(10)$ to $\mathcal{O}(10^4)\,\mathrm{Mpc}$ in comoving size. Consequently, the primordial perturbations sourcing these anisotropies must have been sufficiently small to justify \emph{linear} cosmological perturbation theory on a quasi-de Sitter background as the standard framework for describing the origin of large-scale structure~\cite{Harrison:1967nmu,Guth:1982fnu,Bardeen:1980gcp,Kodama:1984ziu,Mukhanov:1992cpt}.
In this regime, the perturbations are nearly Gaussian, mode coupling is weak, and the background spacetime is well approximated by a rigid \ac{FLRW} metric~---~describing a homogeneous and isotropic Universe on large scales.
Many physically important epochs and mechanisms, however, lie well beyond the linear regime.
Transient departures of the slow-roll phase, ultra-slow-roll phases, sharp features or steps in the inflaton potential, and any other departures from smooth slow-roll dynamics can generate significant non-Gaussianity near horizon crossing, where gradients and constraint equations may play an essential role~\cite{Maldacena:2002vr,Chen:2006xjb,Chen:2008wn,Chen:2011zf,Adshead:2011jq,Martin:2011sn,Chluba:2015bqa,Antony:2022ert,Braglia:2022ftm}.
In these situations, long- and short-wavelength modes interact nonlinearly, and the statistics of the curvature perturbation cannot be captured by linear theory alone.
Analytic tools such as stochastic inflation and Fokker--Planck or Langevin formalisms~\cite{Starobinsky:1986fx,Starobinsky:1994bd} capture the coarse-grained evolution of superhorizon modes and provide important insight into rare fluctuations and tail statistics. However, these formalisms effectively integrate out subhorizon gradients by construction (see also \cite{Habib:1992mu,Stariolo:1994lfp,Tanimura:2006weh,Finelli:2008zg,Finelli:2010sh,Yuste:2016dem,Ezquiaga:2019ftu,Blachier:2023ooc}).
Consequently, they cannot fully describe mode-mode coupling, rescattering, or the backreaction of small-scale inhomogeneities on the near-horizon and long-wavelength dynamics.

A standard approach for such cases is to perform a $(3+1)$-dimensional \emph{lattice simulation}, in which classical fields are evolved in configuration space that automatically takes care of the nonlinear interactions among all populated Fourier modes~\cite{Khlebnikov:1996mc}. 
While 1D linear approximations are restricted by the linearised Fourier equation that depends only on the norm of the wavenumber, a 3D lattice correctly models the phase-space volume available for particle scattering; this is crucial, for instance, for accurately describing thermalisation during preheating~\cite{Felder:2000hr}. 
Furthermore, 3D simulations capture the development of small-scale anisotropies due to the fragmentation of the homogeneous field condensate, which can lead to the generation of sub-Hubble classical gravitational waves~\cite{Khlebnikov:1997di}. 
Another critical regime where such distinctions require careful consideration is in the \acp{PBH} production scenarios~\cite{Zeldovich:1967lct,Hawking:1971ei,Carr:1974nx,Carr:1975qj} (see, \eg, Ref.~\cite{Escriva:2022duf} for a recent review). In these cases, abundances depend exponentially on the extreme tail of the curvature perturbation probability distribution, so that even small deviations from Gaussianity can result in orders-of-magnitude changes in the predicted PBH abundance. 

In its most common implementation, the lattice treats the Universe as a \emph{rigid expanding} FLRW background. In the context of the early Universe, this setup was primarily designed to resolve the growth of sub-Hubble modes during preheating after inflation~\cite{Felder:2000hq,Easther:2010qz,Figueroa:2021yhd,Caravano:2025klk}. 
More recently, the same setup has also been applied to simulate the inflationary dynamics, including axion inflation with an inflaton coupled to gauge fields through a Chern-Simons term~\cite{Caravano:2022epk,Figueroa:2023oxc,Figueroa:2024rkr,Sharma:2024nfu}, and models with a transient \ac{USR} phase~\cite{Caravano:2024moy,Caravano:2024tlp,Caravano:2025diq}.
In such a setup, the fields evolve on a homogeneous background characterised by a single scale factor and expansion rate; inhomogeneities only reside in the matter sector, while the metric carries no spatial structure. 
This works as long as (i) backreaction of inhomogeneities on the expansion is negligible, (ii) curvature sourced by gradients remains subdominant, and (iii) the observables of interest are insensitive to spatial variations in the local Hubble term. 

However, phenomena that motivate real-space simulations~---~such as strong resonance, broad-band mode-mode coupling, stochastic drift of super-Hubble modes, and rare-event tails relevant for PBHs~---~do not by themselves imply the dynamical importance of metric perturbations. Depending on the model and regime, these effects may be described within a rigid-FLRW treatment in which the inhomogeneous gravitational response is neglected. 
Their occurrence, therefore, does not automatically imply that local gravitational effects are dynamically important. 
Nevertheless, during an inflationary Universe where modes continuously cross the Hubble horizon, observables can become sensitive to the local expansion history.
Consequently, it provides physically motivated regimes in which the roles of metric perturbations and inhomogeneous geometry should be assessed on a case-by-case basis. 
When metric perturbations and local gravitational effects do become dynamically relevant, a single, rigid expansion rate cannot capture (a) patch-dependent Hubble friction $3H\dot{\phi}$, (b) contributions of the spatial curvature to the \emph{local} Friedmann constraint, or (c) the correct volume weighting needed to construct uniform-density hypersurfaces for $\delta{N}$ and related observables.
Indeed, although in the typical inflationary models that are conducive to parametric resonance, the influence of local gravity on the small-scale phenomena characteristic of such an epoch is found to be small~\cite{Frolov:2008hy,Huang:2011gf}, in models with a \ac{USR} epoch or an upward turn in the potential~\cite{Kawaguchi:2023mgk}, the importance of inhomogeneous geometry may become enhanced and should therefore be assessed explicitly rather than assumed to be negligible.

In parallel with these developments, fully relativistic simulations of inflationary dynamics and preheating have become increasingly feasible. 
Using \ac{BSSN}-based numerical relativity~\cite{Shibata:1995we,Baumgarte:1998te}, several groups have evolved the full Einstein--scalar (and Einstein-scalar-gauge) system in $3+1$ dimensions, including all tensor, vector, and scalar metric degrees of freedom. Notable applications include investigating the robustness of inflation to generic initial conditions~\cite{East:2015ggf,Clough:2016ymm,Bloomfield:2019rbs,Joana:2020rxm,Aurrekoetxea:2019fhr,Elley:2024alx}, and modeling scalar preheating and oscillon formation~\cite{Aurrekoetxea:2023jwd} in full \ac{GR}, alongside their gravitational-wave signatures and possible black-hole collapse~\cite{Giblin:2019nuv, Kou:2019bbc}.
More recently, fully relativistic studies of gauge preheating~\cite{Adshead:2023mvt} and of the impact of nonlinear gravity on the averaged expansion rate have clarified the regimes where GR qualitatively modifies the dynamics~\cite{Grutkoski:2025ygs}. 
(See Ref.~\cite{Aurrekoetxea:2024ypv} for a recent review on the status of cosmology using numerical relativity.)

Despite such attempts, fully relativistic simulations remain computationally expensive and are typically limited to relatively small volumes, short durations, and narrow regions of parameter space. 
While ideally suited to probing strong-gravity phenomena~---~such as oscillon-dominated phases, black-hole formation, or large-amplitude gravitational waves~---~they are often impractical for broad scans of inflationary models, for systematic studies of rare-event statistics relevant to PBHs, or for large ensembles of realisations. 
In many of these applications, the dominant nonlinear effects are scalar-led, arising from the interplay of inhomogeneous energy density, gradients, and local expansion, while tensor and vector degrees of freedom remain comparatively subdominant.

In this work, we develop an intermediate approach that goes \emph{beyond rigid} FLRW while remaining far less expensive than full numerical relativity. 
We describe the Universe as \emph{inhomogeneous but locally isotropic}, admitting a conformally flat spatial metric with a \emph{local} scale factor $a(\mathbf{x},t)$ and a vanishing shear. 
This `locally FLRW' ansatz retains the computational advantages of standard lattice simulations, while allowing the expansion rate to vary across the grid in a manner consistent with the Hamiltonian and momentum constraints. 
Metric inhomogeneities are encoded in $\psi(\mathbf{x},t)$, which represents the local fluctuation in the number of \efolds relative to the fiducial FLRW background.
Equivalently, $\psi(\mathbf{x},t)$ determines both the local scale factor and the spatial-curvature contribution entering the Friedmann constraint. 
In practice, this framework is designed to improve several aspects of the dynamics and of observable construction:
\begin{enumerate}
    \item Energy-conservation diagnostics, by enforcing a local Friedmann constraint with gradient-induced curvature corrections;
    \item The proper effect of the Hubble friction on the overdense and underdense patches through a spatially varying Hubble rate $H(\mathbf{x},t)$;
    \item The construction of \emph{uniform-density slices} required for $\delta N$ and related statistics, including the correct volume weighting of different regions; and
    \item The sensitivity to non-Gaussian tails, which is controlled by the coupled evolution of local expansion and spatial gradients.
\end{enumerate}

At this point, we note that this approach also distinguishes our scheme from other recent efforts designed to move beyond rigid-FLRW approximations. 
For instance, recent frameworks have incorporated gravitational effects by coupling the nonlinear lattice dynamics of the matter sector to a linearized gravitational sector \cite{Caravano:2024xsb,Jamieson:2025ngu}, typically by solving a linearized constraint equation for the gravitational potential in a specific gauge. 
By contrast, the formalism developed in this work allows us to evolve a local expansion degree of freedom directly at each lattice point through an inhomogeneous FLRW ansatz. 
Rather than expanding the metric perturbations and truncating at linear order, we evolve the local Raychaudhuri equation and retain the nonlinear exponentiation of the local volume,
$a(\mathbf{x},t)=\bar a(t)e^{\psi(\mathbf{x},t)}$.
As long as the shear-free approximation remains valid, this provides a complementary route to incorporate leading gravitational effects beyond a rigid-FLRW background, allowing nonlinear local expansion and proper-volume effects to be treated directly while remaining sufficiently light-weight to explore large volumes, long durations, and broad parameter ranges.
Moreover, because $\psi(\mathbf{x},t)$ directly measures the perturbation in the accumulated local expansion, the same framework provides a natural lattice implementation of the nonlinear $\delta N$ construction on uniform-density hypersurfaces.
It is therefore complementary to full numerical relativity: our scheme is optimised for efficiently resolving nonlinear scalar dynamics and their statistical consequences, whereas fully relativistic simulations remain the gold standard for regimes where anisotropies, tensor modes, or strong-gravity effects become essential.

This paper is organised as follows. In Sec.~\ref{sec:comparison_methods}, we provide a conceptual comparison between our lattice framework and the stochastic-inflation and separate-universe descriptions, clarifying the regimes in which they overlap and differ.
In Sec.~\ref{sec:preliminaries}, we introduce the shear-free, locally FLRW setup used in this work, discuss the relevant averaging and present the corresponding evolution equations, $\delta N$ construction, and statistical estimators employed in our analysis.
In Sec.~\ref{sec:results}, we apply the formalism first to a simple slow-roll model as a consistency check and then to a model with a transient departure from slow-roll, where we study the resulting nonlinear dynamics, curvature perturbations, and non-Gaussian statistics, as well as the regime of validity of the shear-free approximation. 
Finally, we summarise our findings and discuss future directions in Sec.~\ref{sec:conc}.

\section{Conceptual comparison with stochastic inflation}\label{sec:comparison_methods}

Before turning to the explicit equations of motion, it is useful to clarify how the present lattice framework is related to, and differs from, the standard stochastic-inflation formalism. 
Both approaches are designed to describe inflationary fluctuations beyond the simplest linear treatment, but they organise the physics in rather different ways. 
In stochastic formalism for studying inflation, the dynamics are formulated in terms of a coarse-grained, long-wavelength sector sourced by effective noise from shorter modes. 
In contrast, our lattice approach evolves a finite band of Fourier modes directly in real space, including their nonlinear coupling to spatial gradients and to the local expansion rate. 
The purpose of this section is therefore not to argue that the two descriptions are in conflict, but rather to make precise which aspects of the dynamics are naturally captured in each framework and where the lattice provides complementary information. 
This distinction is particularly important when non-linearity becomes significant around the horizon crossing.

Several recent developments have begun to bridge stochastic inflation and real-space nonlinear simulations in complementary ways. 
The authors in Refs.~\cite{Launay:2024qsm,Launay:2025lnc} have studied the stochastic inflation within full general relativity by evolving the infrared modes using the BSSN formulation of numerical relativity, while solving the coarse-grained UV modes via linear perturbation theory.
A different direction is provided by the \ac{STOLAS}~\cite{Mizuguchi:2024kbl,Murata:2026yqb}, which simply implements stochastic inflation in a lattice simulation to capture the nonlinear properties of superhorizon fluctuations within the separate-universe approximation. 
It is cheap and useful to simulate rare events with the use of the importance sampling technique~\cite{Jackson:2022unc}, but it cannot take into account the subhorizon nonlinear dynamics by construction. 

\begin{figure} 
    \centering
    \begin{subfigure}[b]{0.95\textwidth}
        \centering
        \includegraphics[width=\textwidth]{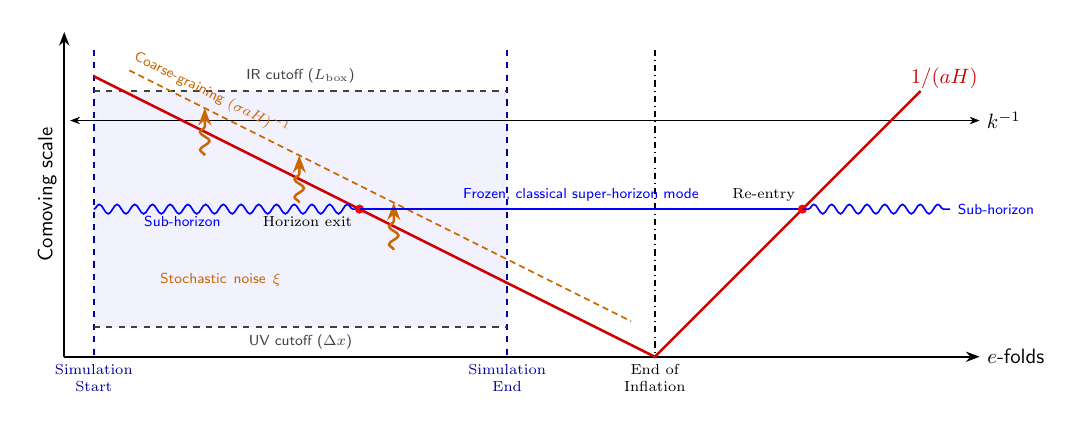}
        \caption{
        This diagram shows the scales and resolution windows for the two methods. In the standard stochastic approach, the field is artificially split by a coarse-graining scale (typically $k_\sigma = \sigma aH$).
        }
        \label{fig:PlotIllsA}
    \end{subfigure}
    
    \vspace{1cm} 
    
    \begin{subfigure}[b]{0.95\textwidth}
        \centering
        \includegraphics[width=\textwidth]{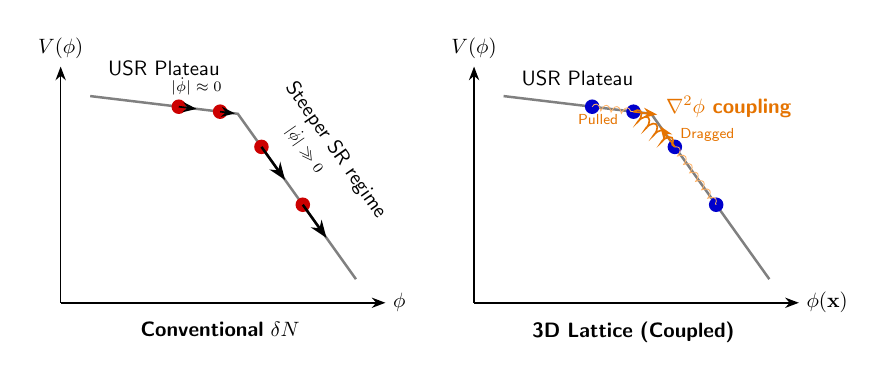}
        \caption{
        Illustration of the different treatment of a sharp potential feature in the conventional $\delta N$ formalism and in the full lattice evolution. 
        \emph{Left}: In the conventional $\delta{N}$ formalism, the IR sector is treated at leading order in the gradient expansion. 
        Each coarse-grained Hubble patch evolves as a locally homogeneous FLRW universe, with no explicit resolved spatial-gradient coupling in the deterministic drift. 
        Two neighbouring patches that cross the kink at slightly different times can therefore acquire very different velocities, leading to a sharp dependence of the local expansion history on the initial condition. 
        \emph{Right}: on the lattice, the scalar field is evolved as a continuous field on a spatial grid. 
        When adjacent cells are around the sharp feature, the resulting mismatch in field values generates a large spatial-gradient coupling that feeds back into the local dynamics. 
        This explicit gradient coupling reduces the velocity contrast across neighbouring cells and spreads the transition over a finite spatial region, yielding a smooth spatial profile of the accumulated number of \efolds for $\delta{N}$, even when the potential itself contains a sharp feature.
        }
        \label{fig:PlotIllsB}
    \end{subfigure}
        \caption{An illustration of how lattice differs in handling the modes from the conventional $\delta{N}$ formalism, based on the leading order of the gradient expansion.
        } 
    \label{fig:PlotIlls}
\end{figure}

In Fig.~\ref{fig:PlotIlls}, we illustrate these dynamical differences between the standard stochastic inflation framework and our full 3D lattice formulation, highlighting how each methodology handles sub-horizon mode evolution in a general simulation, and in handling sharp potential features.
Fig.~\ref{fig:PlotIllsA} shows the relative relation between the comoving scale and \efolds. 
In the stochastic approach, the field is separated by a coarse-graining scale (typically $(\sigma aH)^{-1}$), where short-wavelength quantum modes are integrated out to act as a classical white noise injection, $\xi$, on the super-horizon background.
Conversely, the lattice resolves a finite dynamical window bounded by the infrared ($L$) and ultraviolet ($\Delta x$) cutoffs. 
Rather than utilising idealised noise injection, the lattice explicitly evolves the sub-horizon vacuum fluctuations as they redshift, naturally crossing the shrinking comoving Hubble radius $(aH)^{-1}$ and organically freezing into classical super-horizon perturbations (before eventually re-entering the horizon during the post-inflationary expansion).
This intrinsically bypasses the challenges of formulating non-Markovian colored noise and of capturing the dynamical backreaction between the IR and UV sectors, which currently pose significant theoretical and numerical hurdles in stochastic frameworks 
(for recent advances in this direction, see Refs.~\cite{Mahbub:2022osb,Jackson:2024aoo,Cruces:2024pni,Briaud:2025ayt,Ahmadi:2025oon,Kawasaki:2026hnx}).

In the standard single-field stochastic framework, the long-wavelength IR sector is treated at leading order in the gradient expansion: each coarse-grained Hubble patch evolves as a locally homogeneous FLRW universe, while the shorter-wavelength modes are not evolved explicitly in the IR sector, but are instead encoded in an effective stochastic source for the coarse-grained field. 
Therefore, although the stochastic formalism includes the leading long-wavelength dynamics, it does not retain the explicit resolved spatial-gradient couplings, such as $\nabla^2\phi/a^2$, in the deterministic evolution of the coarse-grained IR field. 
This leading-order gradient treatment of the IR sector is best justified when the coarse-graining scale is chosen well outside the Hubble radius, i.e., for $\sigma\ll 1$, so that the retained long-wavelength modes are deeply super-Hubble.
The trade-off is that a larger set of shorter-wavelength modes is then integrated out, and their nonlinear dynamics around horizon crossing --- together with their feedback on the coarse-grained sector --- must be encoded more carefully in the effective stochastic description. 

The lattice evolution changes this qualitatively, as shown in the right panel. 
On the lattice, the scalar field obeys a partial differential equation containing the resolved spatial-gradient operator, so neighbouring cells remain dynamically coupled. 
When adjacent cells lie on opposite sides of the sharp feature, the resulting mismatch in $\phi$ makes the Laplacian term $\nabla^2\phi/a^2$ large. 
This term reduces the velocity contrast across the kink: the cell that has already started to roll is partially slowed, while the lagging cell is accelerated forward. 
As a result, the transition is spread over a finite spatial region rather than occurring independently point by point. 
The practical consequence is that the lattice produces a smooth, finite $\delta N$ profile without any need to artificially smooth the potential itself. 
This distinction is most relevant for modes near the horizon or coarse-graining scale, where the leading-order gradient expansion begins to lose accuracy while explicit neighbour-to-neighbour coupling remains dynamically important.

\subsection{\label{sec:scales_lattice}Resolved Scales and the Finite Dynamical Range}

A three-dimensional lattice explicitly captures nonlinear spatial coupling, thereby regularising sharp local phase-space transitions.
This comes at the cost of a finite dynamical range.
To clarify the precise kinematic window resolved by our simulations and the regime in which the complementary stochastic formalisms become more efficient, we must first define the comoving scales resolved by our grid.

We evolve the system on a periodic cubic lattice of comoving side length $L$ with $\Ng^3$ grid points, corresponding to a lattice spacing $\Delta x = L/\Ng$. 
The simulation, therefore, resolves a finite band of comoving Fourier modes,
\begin{equation}
  k_{\min} = \frac{2\pi}{L},
  \qquad
  k_{\max} \sim \frac{\pi}{\Delta x},
\end{equation}
with the precise discretized momentum conventions given in Appendix~\ref{app:lattice_details}. 
We choose the box size and initial time so that the modes of interest lie well within the Hubble radius, allowing us to impose adiabatic-vacuum initial conditions as standard in lattice simulations~\cite{Felder:2007les,Figueroa:2020rrl}.
As the universe expands, these comoving modes cross the Hubble horizon and eventually become super-Hubble.
The physical expansion of the background, however, inherently limits the dynamical range of the simulation.
At fixed comoving resolution, the physical lattice spacing grows as $a\Delta{x}$, so the physical UV cutoff redshifts as
\begin{equation}
    k_{\mathrm{phys,max}}(t) \sim \frac{\pi}{a\Delta{x}} \propto a^{-1}.
\end{equation}
Thus, with a fixed comoving grid, the physical UV range progressively shrinks during the evolution. 
Compensating for this loss of physical UV resolution would require refining the lattice as the scale factor grows: the number of grid points per spatial direction would have to increase linearly with the expansion, so that the total number of grid points grows in proportion to the physical volume, i.e., as $a^3$, which is computationally prohibitive. As a result, a lattice simulation can only follow a finite interval of expansion before either the longest modes of interest approach the infrared cutoff set by the box or the shortest relevant modes redshift beyond the resolved ultraviolet range. In practice, this limits the simulation to a finite number of \efolds, with the precise range determined by the box size, resolution, and observables under consideration.


\section{Preliminaries}\label{sec:preliminaries}

In this paper, we address a canonical single-field model of inflation, minimally coupled to gravity, given by the following action
\begin{align}
   S =\int\dd[4]{x} \sqrt{-g}\left[ \frac{\Mpl^2}{2}R - \frac{1}{2}g^{\mu\nu}\partial_{\mu}\phi\partial_{\nu}\phi - V(\phi) \right].
   \label{eq:action}
\end{align}
In this section, we summarise the basic equations and the metric ansatz.

\subsection{Metric ansatz}

We begin from the standard $3+1$ decomposition of the spacetime metric~\cite{Arnowitt:1962hi},
\begin{equation}
    \dd{s^2} = -\mathcal{N}^2 \dd{t^2} + \gamma_{ij}\bigl(\dd{x^i} + \mathcal{N}^i \dd{t}\bigr)\bigl(\dd{x^j} + \mathcal{N}^j \dd{t}\bigr),
    \label{eq:ADMmetric}
\end{equation}
where $\mathcal{N}$ is the lapse function, $\mathcal{N}^i$ is the shift vector, and $\gamma_{ij}$ is the induced metric on constant-time hypersurfaces. 
The inverse metric is given by
\begin{equation}
    g^{00} = -\frac{1}{\mathcal{N}^2}, 
    \qquad
    g^{0i} = \frac{\mathcal{N}^i}{\mathcal{N}^2}, 
    \qquad
    g^{ij} = \gamma^{ij} - \frac{\mathcal{N}^i \mathcal{N}^j}{\mathcal{N}^2},
\end{equation}
with $\gamma^{ij}$ being the inverse of $\gamma_{ij}$.
In this work, we eliminate the lapse perturbation and the shift vector, and write the metric in \emph{synchronous gauge},
$
\mathcal{N} = 1$ and $\mathcal{N}^i = 0$, as
\begin{equation}
    \dd s^2 = -\dd t^2 + \gamma_{ij}(x)\,\dd x^i \dd x^j,
    \label{eq:metric_sync}
\end{equation}
with $x^\mu$ ($\mu=0,\, 1,\, \cdots,\, 3$) being the time and spatial coordinates. As is well known, in this gauge there exist residual gauge degrees of freedom in both the time and spatial coordinates, allowing time-independent coordinate transformations.  We eliminate the former by imposing
\begin{equation}
 \psi(\mathbf{x}, \ti) = 0 \label{Exp:GCini}
\end{equation}
at $t = \ti$. In this work, we ignore the vector and tensor perturbations because vector perturbations decay through the cosmic expansion in a scalar field system and tensor modes (gravitational waves) are assumed to be subdominant to the primary scalar dynamics.

We express the spatial metric as 
\begin{equation}
    \gamma_{ij}(x)
    = a^2(x)\,\tilde{\gamma}_{ij}(x),
    \label{eq:conformal_flat_metric}
\end{equation}
with $\tilde{\gamma}_{ij}$ satisfying $\det \tilde{\gamma}= 1$ and parametrise the local scale factor as
\begin{equation}
    a(x) = \bar a(t)\,e^{\psi(x)}.
    \label{eq:local_scale_factor}
\end{equation}
Here $\bar a(t)$ is a fiducial homogeneous scale factor, which follows the same evolution as the rigid FLRW background, and $\psi(\mathbf{x},t)$ is a nonlinear scalar metric perturbation that encodes local curvature and volume deformations. We express the expansion as $H(x) \equiv \dot{a}(x)/a(x)$, which describes the local isotropic expansion. In this ansatz, the extrinsic curvature is given by 
\begin{equation}
    K_{ij} =  -H \gamma_{ij}+ \sigma_{ij}\,,  \label{Kij}
\end{equation}
whose trace part gives the expansion $K= -3H$ and the traceless part gives the shear, 
\begin{equation}
     \sigma_{ij} \equiv K_{ij} - \tfrac{1}{3}K\,\gamma_{ij} \,. 
\end{equation}
In our gauge, the shear is given by 
\begin{align}
    {\sigma^i}_j = \gamma^{ik} \sigma_{kj} = - \frac{1}{2} \tilde{\gamma}^{ik} \dot{\tilde{\gamma}}_{kj}  \,. 
\end{align}

\subsection{\label{app:volavg}Spatial averaging}

Now, given Eq.~(\ref{eq:local_scale_factor}), the background scale factor $\bar a(t)$ is defined so that the total physical volume of the inhomogeneous slice matches that of the fiducial homogeneous FLRW model. 
Writing $L^3$ for the comoving volume of the periodic box and using $\sqrt{\gamma} = a^3 = \bar a^3 e^{3\psi}$, this can be expressed as
\begin{equation}
\bar a(t)\equiv
\left[
\frac{1}{L^3}\int_{\mathcal D}\dd[3]{x}\,a^3(\mathbf x,t)
\right]^{1/3},
\end{equation}
which is equivalent to the normalisation condition
\begin{equation}
    \big\langle e^{3\psi(\mathbf{x},t)} \big\rangle_{\rm coord} = 1,  \label{Eq:normalization}
\end{equation}
where $\langle\cdots\rangle_{\rm coord}$ denotes a simple average over lattice points.

For any field $X(\mathbf{x},t)$ in the continuum, we define the proper-volume average
\begin{equation}
  \langle X\rangle_V(t)
  \;\equiv\;
  \frac{\displaystyle \int_{\mathcal D} \dd[3]{x} a^3(\mathbf{x},\, t)\,X(\mathbf{x},\, t)}
       {\displaystyle \int_{\mathcal D} \dd[3]{x} a^3(\mathbf{x},\, t)},
  \label{eq:volavg_here}
\end{equation}
On the lattice, this corresponds to
\begin{equation}
  \langle X \rangle_V
  \;=\;
  \frac{\sum_i w_i X_i}{\sum_i w_i}\,,\quad w_i \propto e^{3 \psi(\mathbf{x}_i,\, t)}
  \label{eq:latt_avg}
\end{equation}
where the weights account for the physical-volume expansion of each cell.
For a rigid-FLRW background, one has $\psi = 0$, so all the weights simply reduce to unity, $w_i = 1$.

Because the weights are time-dependent, this averaging operation does \emph{not} commute with time differentiation in general:
\begin{equation}
  \dv{t}\langle X\rangle \neq \langle \dot X\rangle,
\end{equation}
and the resulting covariance terms are precisely the backreaction contributions discussed in Appendix~\ref{app:backreaction}. 
In the conservative regimes studied in this work, these corrections remain numerically small, so the proper-volume averages above provide a consistent notion of effective background quantities. 
For example, the field perturbation is given by
\begin{equation}
\delta\phi(\mathbf{x},t) \equiv \phi(\mathbf{x},t) - \langle\phi\rangle_V(t).
  \label{eq:delphi}
\end{equation}

Now, differentiating the physical volume,
\begin{equation}
  V_{\mathrm{phys}}(t)
  \;\equiv\;
  \int_{\mathcal{D}} \sqrt{\gamma}\dd[3]{x}
  = \bar{a}^3(t)\,L^3,
\end{equation}
with respect to time, we obtain
\begin{equation}
  \frac{\dot{V}_{\rm phys}}{V_{\rm phys}}
  = 3\langle H \rangle_V= 3\,\frac{\dot{\bar a}}{\bar a},
\end{equation}
Hence, the effective background Hubble rate is
\begin{equation}
  \bar H(t) \;\equiv\; \frac{\dot{\bar a}}{\bar a}
  \;=\; \langle H\rangle_V.  
  \label{Eq:Hav}
\end{equation}
That is, once the background scale factor $\bar{a}(t)$ is fixed by matching the total physical volume of the inhomogeneous slice, the corresponding background Hubble rate is determined uniquely: it is the proper-volume average of the local expansion rate.

In our numerical implementation (detailed in Appendix \ref{app:lattice_details}), we evolve $\bar{H}(t)$ via the spatially averaged Raychaudhuri equation and obtain the integrated scale factor $\bar{a}(t)$ (or $\mathbb{N} = \ln \bar{a}$, in practice, see below), solving Eq. (\ref{Eq:Hav}).
The metric perturbation is then reconstructed as
\begin{equation}
  \psi(\mathbf{x},t) = \ln a(\mathbf{x},t) - \ln \bar{a}(t),
\end{equation}
which, by construction, satisfies the normalisation condition (\ref{Eq:normalization}) or, in lattice language, $\frac{1}{\Ng^3}\sum_i e^{3\psi_i(t)} = 1$. 
As our numerical time variable, we use the ``background'' \efolding $\mathbb{N}$ given by $\dd{\mathbb{N}} = \bar{H} \dd{t}$.

\subsection{\label{sec:dec_shear}Equations and decaying shear}

The Hamiltonian and momentum constraints in the $3+1$ form read
\begin{equation}
  {}^{(3)}R + K^2 - K_{ij}K^{ij} = \frac{2}{M_{\rm Pl}^2}\rho, 
\end{equation}
and
\begin{equation}
  D_j K^{j}{}_{i} - D_i K \;=\; -  \frac{1}{M_{\rm Pl}^2}\,\dot{\phi}\,\partial_i\phi,\label{eq:MC_full}
\end{equation}
where ${}^{(3)}R$ is the Ricci scalar of the spatial metric $\gamma_{ij}$ and $D_i$ is the covariant derivative associated with $\gamma_{ij}$. Using Eq.~(\ref{Kij}), one can rewrite the momentum constraints as 
\begin{equation}
  \mathfrak{R}_i(x)
  \;\equiv\;
  \partial_i H
  + \frac{1}{2 \Mpl^2}\,\dot{\phi}\,\partial_i\phi = - \frac{1}{2}D_j \sigma^{j}{}_{i}.
  \label{eq:R_def}
\end{equation}
Since every term in the momentum constraint carries at least one spatial gradient, it trivially vanishes for a homogeneous (background) solution. This aspect requires careful attention in a separate universe approach, in which the entire inhomogeneous Universe is described by gluing numerous homogeneous local universes. 
In such an approach, eliminating redundant degrees of freedom --- for example, the scalar shear --- by solving the momentum constraint introduces non-local operators, such as inverse Laplacians, that obscure the ordering in the gradient expansion.

This difficulty can be avoided for a scalar field system because, at leading order in the gradient expansion, by taking a spatial gradient of the Hamiltonian constraint and using the remaining equations of motion, one recovers the momentum constraint up to an error that decays with the inverse physical volume, i.e., as $1/a^{3}$ in an expanding universe. 
This was shown explicitly under the slow-roll approximation in the context of the $\delta N$ formalism in Ref.~\cite{Sugiyama:2012tj}, and later extended beyond the slow-roll approximation in Ref.~\cite{Garriga:2015tea}.
More generally, an error of the momentum constraints can be shown to fall off as $1/a^{3}$ at the leading order of the gradient expansion under the locality and spatial diffeomorphism invariance, as shown in Ref.~\cite{Tanaka:2021dww}. 
Moreover, the decay of this error is closely related to the decay of the shear itself.
As shown in Ref.~\cite{Tanaka:2021dww}, the spatial diffeomorphism invariance implies that the shear obeys
\begin{equation}
   \frac{1}{H} \dv{(a^3 {\sigma^i}_j)}{t} = {\cal O}\!\left((k/aH)^2\right),
\end{equation}
which implies that the shear decays with $1/a^3$ at leading order in the gradient expansion.
In our simulations, the box is defined in comoving coordinates, so that all modes within the resolved band eventually cross the Hubble radius.
For the scalar-field system considered here, since any sustained anisotropic sources are absent, the shear mode is expected to decay once the relevant modes become super-Hubble.

Considering this behaviour, we adopt a shear-free truncation in the main evolution, while the validity of this truncation becomes subtle,
especially when part of the resolved band remains near or inside the horizon.
Within the decomposition in Eq.~\eqref{eq:conformal_flat_metric}, the condition $\sigma_{ij}=0$ implies that $\tilde{\gamma}_{ij}$ is time independent at leading order in our truncation.
We therefore fix the remaining spatial gauge freedom on the initial slice by imposing $\partial_i \tilde{\gamma}_{ij}=0$, and choose initial data such that $\tilde{\gamma}_{ij}=\delta_{ij}$.
This yields the conformally flat, shear-free spatial metric used in the simulation.
Since some modes are initially sub-Hubble, the validity of this approximation must be checked a posteriori.
As a consistency check, we explicitly monitor the evolution of $\mathfrak{R}_i$, whose decay provides a direct measure of the suppression of the omitted shear sector.
We emphasize that the shear-free ansatz should not be regarded as a truncation whose validity is guaranteed \emph{a priori}. 
Its motivation instead lies in a scale-dependent hierarchy. On sufficiently large (super-Hubble) scales, the omitted shear sector is suppressed by spatial gradients and the decaying shear mode becomes negligible, consistent with the standard gradient-expansion argument. Deep inside the horizon, metric perturbations are generally expected not to provide the leading correction to the matter dynamics as long as they remain perturbative, and the slow-roll hierarchy is not strongly violated. The most delicate regime is around horizon crossing, where neither of these simplifying limits is fully effective, and the validity of the approximation is therefore less automatic.

In our framework, the implicit assumption is that, although spatial gradients are retained to capture the nonlinear evolution of the matter sector, the integrated effect of the neglected shear sector remains subdominant throughout the evolution. For this reason, we do not regard the shear-free approximation as universally valid, but instead assess its consistency on a case-by-case basis through the residual of the momentum constraint. A small residual supports the internal consistency of the approximation for a given simulation, but should not be interpreted as a model-independent proof of validity. In particular, scenarios involving strongly nonlinear metric dynamics or large violations of the slow-roll hierarchy may require a more complete treatment, potentially including the full shear sector or a numerical-relativity formulation.

Under this approximation, the Hamiltonian constraint is given by
\begin{equation}
    H^2
    = \frac{\rho}{3M_{\rm Pl}^2} + \frac{1}{3} C_H,
    \label{eq:HCpFLRW}
\end{equation}
with the energy density of the matter being 
\begin{equation}
    \rho = \frac{1}{2}\dot{\phi}^2 
    + \frac{1}{2a^2}(\vec{\nabla}\phi)^2 
    + V(\phi),
    \label{eq:En_density}
\end{equation}
and $C_H$ being the curvature correction, given by  
\begin{equation}\label{eq: CH}
  C_H
  = \frac{2}{a^2}\left(\nabla^2\psi
  + \frac{1}{2}\bigl(\nabla\psi\bigr)^2\right).
\end{equation}
For our later use, we also define the pressure as:
\begin{align}
  p
  = \frac12 \dot\phi^2
   - \frac{1}{6a^2}\big(\vec{\nabla}\phi\big)^2
   - V\bigl(\phi\bigr).
   \label{eq:pressure}
\end{align}
Here, the spatial gradients are taken with respect to the comoving coordinates, e.g., $\nabla^2 = \delta^{ij}\partial_i\partial_j$.
Equation~\eqref{eq:HCpFLRW} is nonlinear in $\psi$ and enforces the local Friedmann relation between the inhomogeneous expansion rate, the scalar field energy density, and the spatial curvature induced by $\psi$. The scalar field obeys the \ac{KG} equation, given by 
\begin{equation}
    \ddot{\phi} + 3H\dot{\phi}
    - \frac{1}{a^2}\left(\nabla^2\phi 
    + \nabla\phi\cdot\nabla\psi\right)
    + V_{\phi} = 0,
    \label{eq:KGpFLRW}
\end{equation}
where $V_{\phi} \equiv \partial V/\partial\phi$.

\subsection[\texorpdfstring{$\delta N$}{delN} formalism on lattice]{\label{sec:deltaN_formalism}\boldmath \texorpdfstring{$\delta N$}{deltaN} formalism on lattice}

The $\delta N$ formalism \cite{Starobinsky:1982ee, Starobinsky:1986fxa, Sasaki:1995aw, Sasaki:1998ug, Lyth:2004gb} was developed to calculate the primordial spectrum of the adiabatic curvature perturbation originally based on the gradient expansion. 
The $\delta N$ formalism has been widely used in studies of the early universe, as it can simplify the computation of superhorizon dynamics and provide an intuitive understanding of the evolution of primordial fluctuations. Recently, the $\delta N$ formalism was generalized to compute arbitrary scalar, vector, and tensor type perturbations, including GWs sourced by arbitrary bosonic fields, dubbed the generalized $\delta N$ formalism (g$\delta N$ formalism) \cite{Tanaka:2021dww, Tanaka:2023gul, Tanaka:2024mzw}. 
In what follows, we develop a lattice implementation of the $\delta{N}$ formalism within our locally FLRW framework, allowing the effect of subhorizon inhomogeneities on the local expansion to be treated directly.

The local number of \efolds of expansion between two time slices, $t_i$ and $t_f$, is given by
\begin{align}
N(\mathbf{x};t_i,t_f) = \int_{\ti}^{\tf}H(\mathbf{x},t)\dd{t},
\end{align}
and defining $\mathbb{N} \equiv \log{\bar{a}}$ or, equivalently, $\dd{\mathbb{N}} = \bar{H} \dd{t}$ (notice that, in general, $\langle N(\mathbf{x},t)\rangle_{\rm coord} \neq \mathbb{N}(t)$), the perturbation in the local expansion is naturally defined by 
\begin{align}
    \delta N(\mathbf{x};t_i,t_f) \equiv N(\mathbf{x};t_i,t_f) - \mathbb{N}(t_i,t_f) =\psi(\mathbf{x}, \tf) - \psi(\mathbf{x}, \ti) \,. 
\end{align}
Because the residual gauge freedom is fixed on the initial slice by imposing $\psi(\mathbf{x},t_i)=0$ (cf. Eq.~(\ref{Exp:GCini})), the initial slice contribution in $\delta N$ vanishes.
We determine the final slice $t=t_f$ by the condition that the local energy density reaches a prescribed value $\rho=\rho_f$,
\begin{equation}
  \rho(\mathbf{x},t_f(\mathbf{x})) = \rho_f.
\end{equation}
In the chosen synchronous gauge, the corresponding coordinate time $t_f(\mathbf{x})$ is spatially dependent: different lattice sites reach the same density at different times.
Since the degree of freedom to choose the time coordinate is already exhausted, we cannot further choose an additional gauge transformation to uniform-density slicing; rather, we identify the uniform-density hypersurface directly within the fixed synchronous coordinate system.
The resulting nonlinear perturbation in \efolds is therefore
\begin{equation}
  \delta N(\mathbf{x})
  = \psi(\mathbf{x},t_f).
\end{equation}
In general, this nonlinear geometric $\delta N(\mathbf{x})$ need not have vanishing spatial average. 
For comparison with the fluctuation field used in observational spectra, we therefore define
\begin{equation}
  \zeta(\mathbf{x},t_f)
  \equiv
  \delta N(\mathbf{x})-\langle\delta N(\mathbf{x})\rangle_{\mathrm{coord}}
  =
  \psi(\mathbf{x},t_f)-\langle\psi(\mathbf{x},t_f)\rangle_{\mathrm{coord}},
\end{equation}
whose coordinate-space average vanishes by construction. 
For the periodic lattice, this means that the spectra of $\delta N$, $\zeta$, and any complementary estimator differing only by a homogeneous offset coincide for all physical modes $k>0$, once the zero mode is treated consistently.

\subsubsection{Computation protocol}

The above $\delta{N}$ can be computed by solving a local expansion equation,
\begin{equation}
  \dv{\ln a(x)}{t} = H(x),
\end{equation}
starting from an initial flat slice and asking \emph{how much local expansion is accrued by the time we hit the chosen final condition at each point?}
In terms of the background \efolds coordinate $\mathbb{N}$, this becomes
\begin{equation}
  \dv{\psi(\mathbf{x},\mathbb{N})}{\mathbb{N}} = \frac{H(\mathbf{x},\mathbb{N}) }{\bar{H}} -1,  \label{Eq:Psi}
\end{equation}
which is precisely the evolution equation used for the local expansion field in the full inhomogeneous runs. 

As a complementary estimator, we also compute the quantity
\begin{equation}
  \delta N_\rho(\mathbf{x})
   \equiv \mathbb{N}(t_f(\mathbf{x}))
   - \big\langle \mathbb{N}(t_f(\mathbf{x}))\big\rangle_{\mathrm{coord}},
\end{equation}
which is obtained directly from the background e-fold clock $\mathbb{N}$ evaluated at the local hitting time $t_f(\mathbf{x})$.
While $\mathbb{N}$ itself is defined from the averaged Hubble rate, its spatial dependence enters through the non-uniform final time $t_f(\mathbf{x})$.
When the leading-order gradient expansion provides a good approximation throughout the evolution, this quantity reproduces the same spatially varying part of the nonlinear $\zeta$ obtained from Eq.~\eqref{Eq:Psi}, differing at most by a homogeneous offset. Since, on a periodic lattice, subtracting the spatial mean affects only the $k=0$ mode, the spectra of $\delta N_\rho$ and the geometric nonlinear $\zeta$ coincide for all physical $k>0$ modes once the zero mode is removed.
Similarly, one may define a $\delta N_\phi$ estimator by replacing the final $\rho=\rho_f$ slicing with a fixed-$\phi$ slicing.

\subsubsection{\label{sec:liner_zetas}Estimators of time evolution}

Although the nonlinear curvature perturbation, such as using $\delta N_\rho$ (or $\delta N_\phi$), can in principle be constructed on any chosen target slicing during the simulation, their interpretation as large-scale curvature perturbations is clearest when the relevant modes are super-Hubble, and the separate-universe approximation is applicable.
For this reason, we measure these nonlinear quantities at the final instant of the simulation, when all the modes are super-Hubble.
Additionally, we introduce linearised estimators that can be evaluated directly at each (synchronous) time step and used to continuously track the intermediate evolution.
We define the so-called curvature perturbation on the uniform-density slice by~\cite{Bardeen:1983qw,Martin:1997zd}
\begin{equation}
  \zeta^{\mathrm{est}}(x)
  = \psi(x)
    - \bar H\,\frac{\delta\rho(x)}{\dot{\bar\rho}},
  \label{eq:zeta_lin}
\end{equation}
and for the uniform-field (comoving) slicing~\cite{Liddle:1993fq,Lyth:1998xn}
\begin{equation}
  \mathcal{R}^{\mathrm{est}}(x)
  = \psi(x)
    - \bar H\,\frac{\delta\phi(x)}{\dot{\bar\phi}},
  \label{eq:R_c}
\end{equation}
where
$
\delta\rho(x)\equiv \rho(x)-\bar\rho
$
and
$
\delta\phi(x)\equiv \phi(x)-\bar\phi
$.
These quantities are computed from the nonlinear lattice fields at each time step. 
We use the superscript `$\mathrm{est}$' to distinguish them from the fully nonlinear quantities obtained from the $\delta N$ construction on the final slicing.
Here $\bar\rho$, $\bar p$, and $\bar\phi$ denote the proper-volume averages defined above, while $\dot{\bar\rho}$ and $\dot{\bar\phi}$ are the time derivatives of these averaged background quantities.\footnote{Under the time coordinate transformation $t \to \tilde{t}= t + \delta t$, a scalar quantity, $\rho$ (or $\phi$), transforms as 
\begin{equation}
    \tilde{\rho}(\mathbf{x},\, t) = \rho(\mathbf{x},\, t) - \dot{\rho} (\mathbf{x},\, t) \delta t + \cdots. 
    \label{eq:rho_tilde}
\end{equation}
Inserting $\rho = \bar{\rho}  + \delta \rho$ and $\tilde{\rho} = \bar{\rho} + \delta \tilde{\rho}$ into this expression, one obtains $\delta \tilde{\rho} = \delta \rho - \dot{\bar{\rho}}\, \delta t$ at linear order in perturbation. Since the gauge invariant variable is constructed by using this expression, we use $\dot{\bar{\rho}}$ instead of $\bar{\dot{\rho}}$ in the definition of $\zeta^{\mathrm{est}}$. The same story also follows for ${\cal R}^{\mathrm{est}}$.} 
Since proper-volume averaging and time differentiation do not commute in general, they differ from the averages of the local time derivatives $\overline{\dot\rho}\equiv\langle\dot\rho\rangle$ and $\overline{\dot\phi}\equiv\langle\dot\phi\rangle$, respectively. 
The distinction is controlled by the covariance terms discussed in Appendix~\ref{app:backreaction}. In the conservative regimes studied here, these corrections remain numerically small, so that $\dot{\bar\phi}\approx \overline{\dot\phi}$, and $\dot{\bar\rho}$ is well approximated by the effective background evolution adopted in the code.

The estimators $\zeta^{\mathrm{est}}$ and $\mathcal{R}^{\mathrm{est}}$ coincide with the corresponding gauge-invariant curvature perturbations only at linear order.
Nevertheless, they provide convenient diagnostics of the time evolution and are useful for tracking the approach to the final conserved super-horizon curvature perturbation.
In a single-field model, the two coincide on super-horizon scales once the system has returned to an adiabatic attractor.

For our later use, we also define the power spectrum $P_X(k)$ for any statistically homogeneous and isotropic field $X(\mathbf{x})$ as
\begin{equation}
\langle{X_{\mathbf{k}}X_{\mathbf{k'}}^{\ast}}\rangle = (2\pi)^3\delta^{(3)}(\mathbf{k} - \mathbf{k}')P_X(k),
    \label{eq:PX}
\end{equation}
where $X_{\mathbf{k}}$ is the Fourier transform of $X(\mathbf{x})$. 
It is worth noting that, as in our locally FLRW ansatz, we dynamically evolve the spatial metric field $\psi(x)$, both the physical volume element and the local physical momenta vary across the lattice.
Consequently, we will define the spectra in terms of coordinate (comoving) power spectra evaluated with respect to the background spatial grid, following the standard practice for extracting the asymptotic gauge-invariant observables such as the $\delta N$ power spectrum~\cite{Salopek:1990jq,Sasaki:1995aw,Lyth:2004gb}. 
The dimensionless power spectrum is correspondingly defined as
\begin{equation}
    \Delta_X^2(k) \equiv \frac{k^3}{2\pi^2}P_X(k),
\end{equation}
while $k$ is identified with effective lattice modes $k_{\mathrm{eff}}$ (see Appendix~\ref{app:init_kmodes} for definition) to match the continuous spectra~\cite{Stamatopoulos:2012np,Caravano:2021pgc,Caravano:2022epk}. 

\subsection{\label{subsec:NG_measures}Non-Gaussianity measures}

In order to quantify the non-Gaussian statistics of the curvature perturbation, we work with the one-point probability distribution function
of the curvature perturbation $\zeta^{\mathrm{est}}$ reconstructed on the lattice at the final uniform-density (or uniform-$\phi$) slice, as well as its linear estimators (which are computed on equal background \efolds).
Now, the central moments of $\zeta^{\mathrm{est}}$ are defined similarly as
\begin{equation}
\mu_n \equiv 
  \frac{\sum_i w_i \,(\zeta^{\mathrm{est}}_i - \bar\zeta^{\mathrm{est}})^n}{\sum_i w_i},
  \qquad n=2,3,4,\dots;
  \qquad
  \bar\zeta^{\mathrm{est}} \equiv \langle \zeta^{\mathrm{est}} \rangle\,.
\end{equation}
In our simulations, we explicitly subtract the measured box mean $\bar\zeta$ before computing moments, so that $\mu_1 = 0$ by construction even in a finite volume.
The variance is $\sigma^2 \equiv \mu_2$, and we use the fourth cumulant,
\begin{equation}
\kappa_4 \equiv \mu_4 - 3\mu_2^2,
\end{equation}
so that a purely Gaussian distribution has $\mu_3 = 0$ and $\kappa_4 = 0$~\cite{peebles2020large}.
We also compute the reduced (hierarchical) skewness and kurtosis as prevalent in cosmology~\cite{peebles2020large,Bouchet:1992ApJ,Juszkiewicz:1993uw,Bernardeau:1993qu}:
\begin{align}
S_3 \equiv \frac{\mu_3}{\mu_2^2}
 = \frac{\mu_3}{\sigma^4};\quad
 S_4 \equiv \frac{\kappa_4}{\mu_2^3}
        = \frac{\mu_4 - 3\mu_2^2}{\mu_2^3}.
 \label{eq:S3_S4_red_def}
\end{align}
In contrast to the standard $S_3^{\mathrm{std}} \equiv \mu_3/\sigma^3 = \mu_3/\mu_2^{3/2}$ used in statistics, the reduced $S_3$ is more convenient for comparisons with local-type non-Gaussianity as it scales as the inverse powers of the variance and remains finite in the small-variance limit.

To interpret these moments in terms of the usual local-type non-Gaussianity parameters $(f_{\rm NL}, g_{\rm NL})$, we consider the standard local ansatz for the curvature perturbation,
\begin{equation}
  \zeta
  \;=\;
  \zeta_g
  + \frac{3}{5} f_{\rm NL}
    \Bigl[\zeta^2_g - \langle \zeta_g^2 \rangle\Bigr]
  + \frac{9}{25} g_{\rm NL}\,\zeta^3_g
  + \cdots,
  \label{eq:local-ansatz}
\end{equation}
where $\zeta_g$ is a Gaussian random field with zero mean and variance $\langle \zeta_g^2 \rangle = \sigma_g^2$. The subtraction of $\langle \zeta_g^2 \rangle$ in the quadratic term ensures that $\langle \zeta \rangle = 0$ at leading order. 
In this convention, the connected moments of $\zeta$ can be expressed perturbatively in powers of $(f_{\rm NL}, g_{\rm NL})$.
For small non-Gaussianity, one finds, to leading order in $f_{\rm NL}$ and $g_{\rm NL}$,
\begin{align}
  \mu_2 &\simeq \sigma_g^2, \\
  \mu_3 &\simeq
    \frac{18}{5} f_{\rm NL}\,\sigma_g^4, \\
  \mu_4 - 3\mu_2^2 &\simeq
    \frac{216}{25}\left(g_{\rm NL} + 2 f_{\rm NL}^2\right)\sigma_g^6,
\end{align}
which, using the relations in Eq.~\eqref{eq:S3_S4_red_def}, can be written as:
\begin{equation}
  S_3 \;\equiv\; \frac{\mu_3}{\mu_2^2}
      \;\simeq\; \frac{18}{5} f_{\rm NL},
  \qquad
  S_4 \;\equiv\; \frac{\mu_4 - 3\mu_2^2}{\mu_2^3}
      \;\simeq\; \frac{216}{25}\left(g_{\rm NL} + 2 f_{\rm NL}^2\right).
\end{equation}
Finally, inverting these relations yields
\begin{align}
  f_{\rm NL}^{\rm(1pt)}
    &\;\equiv\; \frac{5}{18}\,S_3,\label{eq:fNL_from_S3} \\
  g_{\rm NL}^{\rm(1pt)}
    &\;\equiv\; \frac{25}{216}\,S_4 - 2\bigl(f_{\rm NL}^{\rm(1pt)}\bigr)^2.\label{eq:gNL_from_S3}
\end{align}
We use the superscript `$(1\mathrm{pt})$' to emphasise that these are \emph{effective} non-linearity parameters inferred from the one-point distribution under a local ansatz assumption. 
When the dynamics produces a non-local shape or strongly scale-dependent non-Gaussianity, the above mapping should be regarded as a diagnostic rather than a literal $f_{\rm NL}$ or $g_{\rm NL}$ in the bispectrum/trispectrum sense.

In practice, for a given lattice realisation, we estimate the moments $\mu_n$ using the weighted averages defined above and compute the reduced skewness and kurtosis $(S_3, S_4)$, together with the associated one-point nonlinearity parameters $(f_{\rm NL}^{\rm(1pt)}, g_{\rm NL}^{\rm(1pt)})$. These quantities are evaluated from the grid values of $\zeta^\est$ as defined in Sec.~\ref{sec:liner_zetas}, allowing us to track their time evolution throughout the simulation. 
We also evaluate the same statistics at the final time using $\zeta = \delta N$ defined on an equal-density (or constant-$\phi$) hypersurface. 

These one-point measures provide a compact characterisation of the strength of non-Gaussian tails in the distribution of curvature perturbation.
This characterisation can be particularly useful in regimes where the non-Gaussianity is induced by a sudden departure from the slow-roll phase.
They serve as a complement to a full bispectrum-based analysis, which we defer to future work.

\section{Results}\label{sec:results}

In this section, we describe the main numerical results obtained from the inhomogeneous
lattice simulations described in Appendix~\ref{app:lattice_details}.
The initial field values for the simulation are determined by using the linear mode function for the adiabatic vacuum, as detailed in Appendix~\ref{app:ICs}.
In addition to the main dynamical variables, we monitor the usual Hubble slow-roll parameters, including gradient contributions. 
In analogy with the homogeneous FLRW case, the effective Hubble slow-roll parameter is computed as
\begin{equation}
  \epsilon_H
  \;=\; \frac{3}{2}\,\frac{\bar\rho+\bar p}{\bar\rho}
  \;=\; \frac{3}{2}\frac{\displaystyle
          \big\langle \dot\phi^2
          + \tfrac{1}{3a^2}\big(\vec{\nabla}\phi\big)^2 \big\rangle}
         {\displaystyle
          \big\langle \tfrac12 \dot\phi^2
          + \tfrac{1}{2a^2}\big(\vec{\nabla}\phi\big)^2
          + V(\phi)\big\rangle}\,,
  \label{eq:epsilonH_def}
\end{equation}
so that gradient energy contributes on the same footing as the homogeneous kinetic term.
Here we have used $\bar\rho \equiv \langle\rho\rangle$ and $\bar p  \equiv \langle p\rangle$ (cf. Eqs.~(\ref{eq:En_density}) and (\ref{eq:pressure})).
In the exact FLRW limit, this coincides with 
$
\epsilon_H\;\equiv\; -\dot{\bar{H}}/\bar{H}^2
$.
Using the effective $\epsilon_H$ above, the second Hubble slow-roll parameter is defined in the standard way,
\begin{equation}
  \eta_H \;\equiv\; \dv{\ln\epsilon_H}{\mathbb{N}}
  \;=\; \frac{1}{\epsilon_H}\,\frac{1}{\bar H}\,\dv{\epsilon_H}{t},
  \label{eq:etaH_def}
\end{equation}
and is similarly constructed from the lattice-averaged quantities at each time slice.
In the strictly homogeneous limit ($\nabla\phi\rightarrow 0$), they reduce to the familiar single-field expressions; in the fully inhomogeneous case, they quantify the effective departure from the slow-roll due to additional pressure and energy density contributions from evolving field fluctuations.

With these basic definitions in mind, we now present the results of our lattice simulation for the two broad classes of models: 
\begin{itemize}
  \item \emph{Slow-roll inflation models} 
  
    These are models in which the background trajectory remains close to the slow-roll attractor throughout the evolution. 
    In this class, the inflaton evolves on a sufficiently smooth potential without violating the slow-roll condition, and the dynamics remain close to the standard quasi-de Sitter picture.
    
  \item \emph{Non-slow-roll inflation models}
  
  These are models in which the inflaton temporarily departs from the slow-roll attractor. 
  In the present work, our main example is a model with an intermediate \ac{USR} phase~\cite{Starobinsky:1992ts} embedded between two slow-roll stages. 
  More generally, this class can display a wide variety of behaviours depending on the duration of the non-slow-roll phase, the sharpness of the transition, and the detailed morphology of the scalar potential~\cite{Kawaguchi:2023mgk}. 
  For this reason, the detailed phenomenology discussed below should be understood as a representative of the specific examples considered here.
\end{itemize}
We first present the results from a slow-roll model, where the lattice mainly serves as a consistency check against the standard perturbative picture. 
We then turn to the non-slow-roll models, where departures from the slow-roll attractor lead to qualitatively new effects in the evolution of curvature perturbations and their statistics.

\subsection{\label{sec:sr_model}A warm-up for simple slow-roll model}

To validate the code and to establish notation and conventions, we first present the results for a simple slow-roll inflationary model. 
The methods employed and the qualitative features discussed here do not depend on the specific form of the inflationary potential and therefore serve as a benchmark for the more general, non-slow-roll scenarios considered later.
As a concrete example, we consider the quadratic inflationary potential
$
  V(\phi) = \tfrac{1}{2} m^2 \phi^2,
$
which serves as a prototypical model of chaotic inflation~\cite {Linde:1983gd,Linde:1984ir}.
Although now observationally disfavoured, its simplicity and well-understood linear dynamics make it a standard benchmark for validating lattice simulations of the early Universe~\cite{Caravano:2021pgc,Caravano:2022epk,Figueroa:2023oxc,Figueroa:2024rkr,Sharma:2024nfu}.

We start the simulation with $\phi_{\mathrm{in}} = 14.5M_{\mathrm{Pl}}$, and we normalise the scale factor to $a = 1$ on the initial slice.  With this choice, modes that exit the horizon roughly $4.5$ \efolds after the start of the simulation correspond to those that exit about $50$ \efolds before the end of inflation in the background solution.  We take the mass parameter $m = 7.5 \times 10^{-6}M_{\mathrm{Pl}}$, which yields a scalar curvature power spectrum amplitude of $\Delta^2_\mathcal{R} \simeq 2.19 \times
10^{-9}$ at CMB scales~\cite{WMAP:2003ivt,Planck:2018vyg}. 

The evolution of the different energy components is shown in the right panel of Fig.~\ref{fig:figBG1Chaotic}. 
In addition to the kinetic ($\rho_K$), gradient ($\rho_G$), and potential ($\rho_P$) contributions defined in Eq.~\eqref{eq:En_density}, we also track the `curvature' term, $C_H$~\eqref{eq: CH}, 
where the angular brackets in Fig.~\ref{fig:figBG1Chaotic} denote the volume-weighted lattice averages introduced in Eq.~\eqref{eq:latt_avg}.
As expected, the contribution from $C_H$ is extremely small throughout the evolution. At the initial slice ($\bar{a} = 1$), we impose $\psi = 0$, so $C_H$ vanishes by construction and then exhibits a rapid rise for $t>0$ as soon as nonzero gradients of $\psi$ are generated dynamically.
The gradient contribution from the scalar field is also very small.
\begin{figure} 
    \begin{center}
    \begin{minipage}{0.44\textwidth}
        \centering
        \includegraphics[width=\linewidth,height=0.85\linewidth,keepaspectratio]{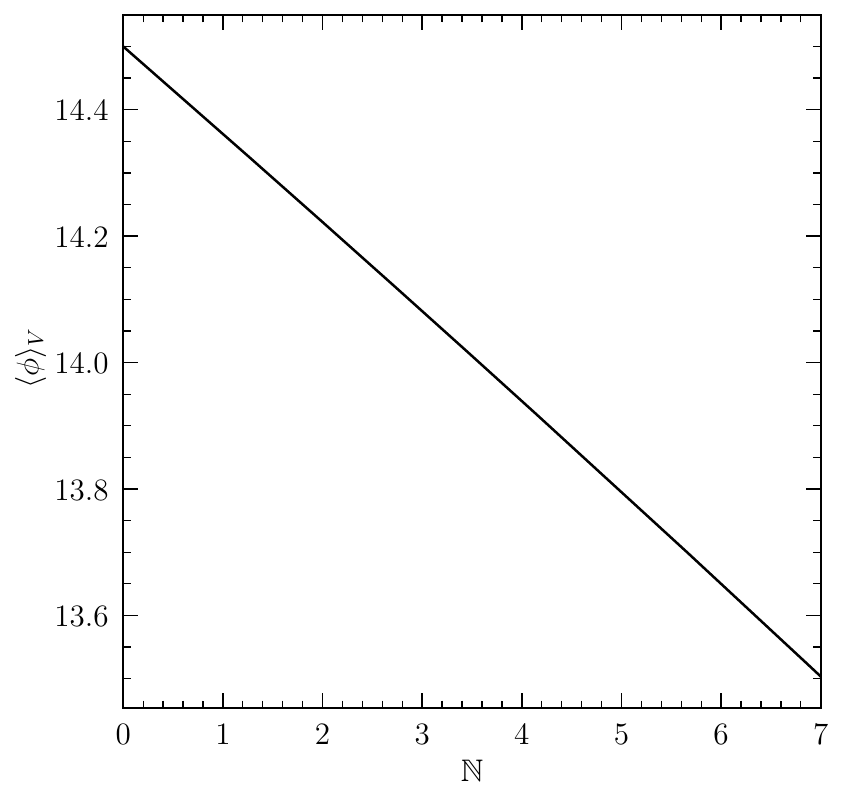} 
    \end{minipage}
    \begin{minipage}{0.44\textwidth}
        \centering
        \includegraphics[width=\linewidth,height=0.85\linewidth,keepaspectratio]{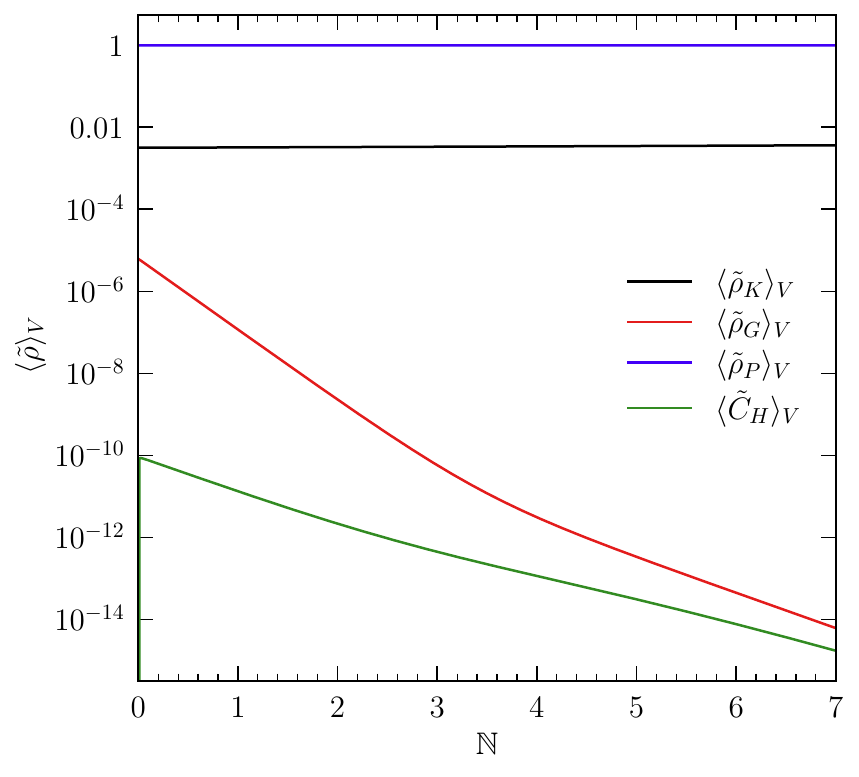} 
    \end{minipage}
    \end{center}
    \caption{
    We plot the evolution of the volume-averaged field value (left panel) and the energy components (normalised by the total energy $\langle{\tilde{\rho}_i}\rangle_V = \langle{\rho_i}\rangle_V/\langle{\rho_{\mathrm{tot}}}\rangle_V$) (right panel) for the
    $m^2\phi^2$ model.  
    }
    \label{fig:figBG1Chaotic}
\end{figure} 
Both the gradient and curvature pieces carry an explicit factor $a^{-2}$, but their \emph{volume-averaged} amplitudes need not decay at the same rate because they probe different combinations of fields and Fourier modes: schematically,
$
\big\langle (\nabla\phi)^2 \big\rangle \sim \int \dd[3]{k} k^2 |\phi_k|^2,
$
while
$
\big\langle \nabla^2\psi \big\rangle \sim \int \dd[3]{k} k^2 \psi_k
$
and
$
\big\langle (\nabla\psi)^2 \big\rangle \sim \int \dd[3]{k} k^2 |\psi_k|^2.
$
When some modes are still inside the horizon, $\phi_k$ and $\psi_k$ obey different equations of motion, and their gradients can redshift with distinct effective rates.
Only once all relevant modes are safely super-horizon and approximately frozen in comoving coordinates do both gradient and curvature contributions track the trivial
$a^{-2}$ scaling associated with the expanding background.
\begin{figure}
    \begin{center}
    \begin{minipage}{0.44\textwidth}
        \centering
        \includegraphics[width=0.9\textwidth]{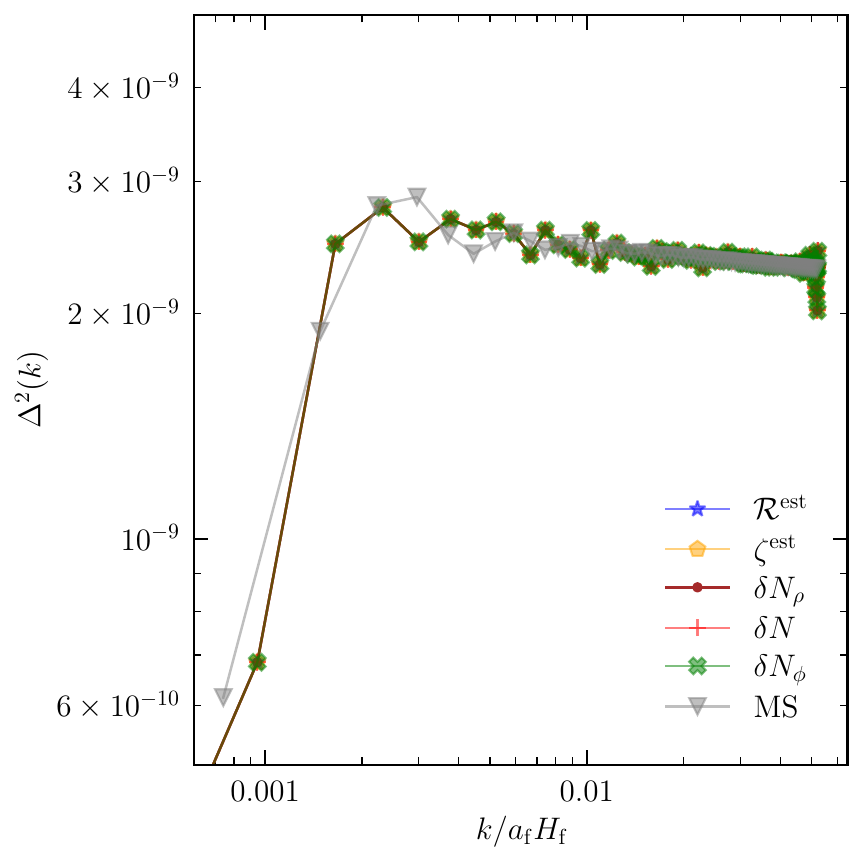} 
    \end{minipage}
    \begin{minipage}{0.44\textwidth}
        \centering
        \includegraphics[width=0.9\textwidth]{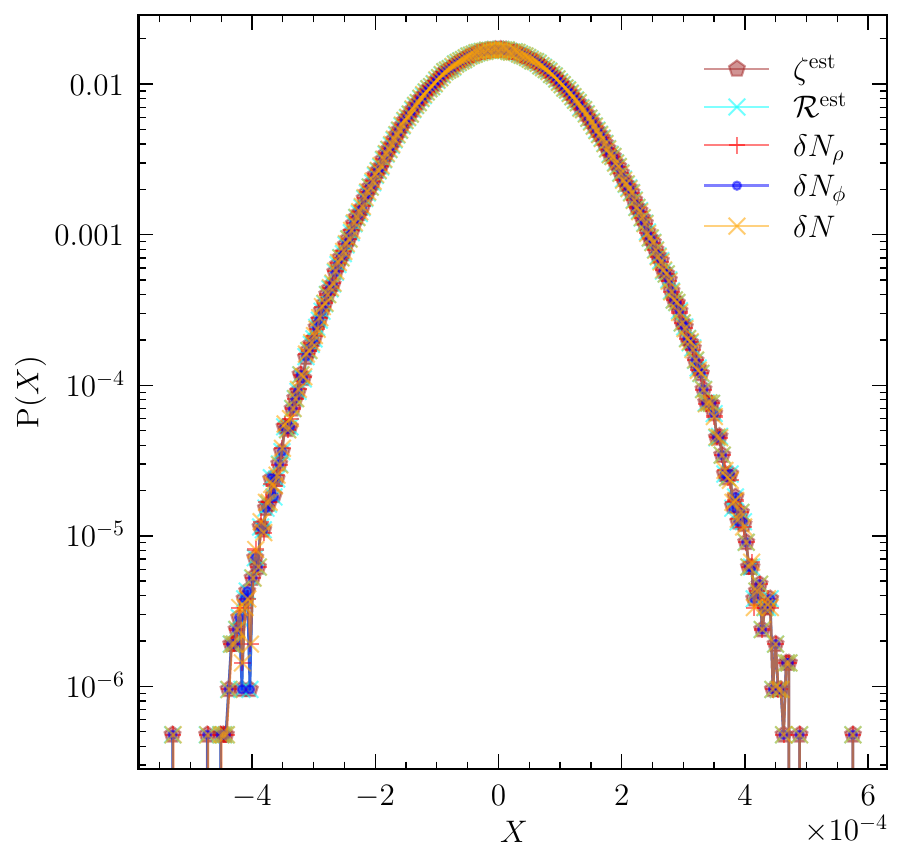} 
    \end{minipage}
    \end{center}
    \caption{
    These plots show the final spectra and \acp{PDF} of different variables for the $m^2\phi^2$ model. 
    }
    \label{fig:figSpectraChaotic}
\end{figure} 
In this slow-roll setup, we find that the lattice spectra of the gauge-invariant
curvature perturbations $\zeta^{\mathrm{est}}$ and $\mathcal{R}^{\mathrm{est}}$, as well as the $\delta{N}$ curvature obtained from our separate-universe and constant-density constructions, all agree with linear-theory predictions within numerical accuracy, as shown in Fig.~\ref{fig:figSpectraChaotic}.
This case thus provides a stringent baseline test for the more nontrivial, non-slow-roll examples studied in the following sections.
It is also worth noting that our spectra are remarkably consistent with those obtained from the \ac{MS} equation, which fully incorporates the scalar metric perturbations at the linear level, whereas the \ac{KG} equation in the rigid \ac{FLRW} is known to deviate from the \ac{MS} result. 
This suggests that the local \ac{FLRW} approximation captures the leading gravitational corrections controlling the scalar power spectrum.

\subsection{\label{sec:nsr_model}A model for beyond slow-roll dynamics}
We next take up a class of single-field models in which the inflaton evolves along a piecewise-linear potential with one or more sharp changes in slope. The prototype of this family is the Starobinsky's linear model~\cite{Starobinsky:1992ts}, in which the inflaton rolls on a linear potential that undergoes an abrupt break at some field value $\phi_1$.  
Far from the break, the dynamics are well described by ordinary slow-roll inflation; near the break, the sudden change in $V_{\phi}$ violates the slow-roll conditions for a brief period, and imprints localised features in the curvature power spectrum.  Depending on the sign and magnitude of the slope change, the background can transiently enter an \ac{USR} regime~\cite{Martin:2011sn,Pi:2022zxs} or exhibit more general transient behaviour, and smoothing the discontinuity to a finite-width step leads to related phenomenology with broadened features~\cite{Kawaguchi:2023mgk}.

In this work, we use a slightly generalised version of the original Starobinsky's linear potential,
allowing for two successive kinks in the slope:
\begin{align}
  V(\phi) = 
  \begin{cases}
    V_0 + v_1(\phi - \phi_1) & \text{for } \phi > \phi_1,\\
    V_0 + v_2(\phi - \phi_1) & \text{for } \phi_1 \geq \phi \geq \phi_2,\\
    V_0 + v_2(\phi_2 - \phi_1) + v_3(\phi - \phi_2) & \text{for } \phi < \phi_2,
  \end{cases}
  \label{eq:star3}
\end{align}
with $V_0$ setting the overall energy scale, and $v_1$, $v_2$, and $v_3$ the slopes in the three regions. 
The additive constant in the third branch is chosen such that the potential is continuous at $\phi_2$.  The break points $\phi_1$ and $\phi_2$ mark field values where the slope (and hence the slow-roll parameter $\epsilon_V=\frac{\Mpl^2}{2}\pqty{\frac{V_\phi}{V}}^2$) changes abruptly. 
The parameters $v_1$ and $v_2$ set the slopes in the two regions, which we parameterised relative to the slope in the first region by setting:
$v_2 = \frac{v_1}{\Lambda_1}$ and $v_3 = \frac{v_1}{\Lambda_2}$.
For $\Lambda_1 > 1$, the inflaton first rolls down a steeper linear segment with slope $v_1$ for $\phi > \phi_1$, and then encounters a transition at $\phi = \phi_1$ to a flatter linear tail with slope $v_1/\Lambda_1$. 
The second break at $\phi_2$ then marks the onset of the region with a slope $v_1/\Lambda_2$.
The potential itself is continuous by construction at both $\phi_1$ and $\phi_2$, while $V_{\phi}$ exhibits two sharp kinks, whose sizes and signs are controlled by $\Lambda_1$ and $\Lambda_2$. 
The original Starobinsky's linear model, with a single localised change in slope, is recovered as the special case $\Lambda_1 = \Lambda_2$, in which the segments for $\phi_1 \ge \phi \ge \phi_2$ and $\phi < \phi_2$ share the same slope and the only genuine kink in $V_{\phi}$ occurs at $\phi_1$.

\begin{figure} 
    \centering
    \begin{minipage}{0.49\textwidth}
        \centering
        \includegraphics[width=\linewidth,height=0.85\linewidth,keepaspectratio]{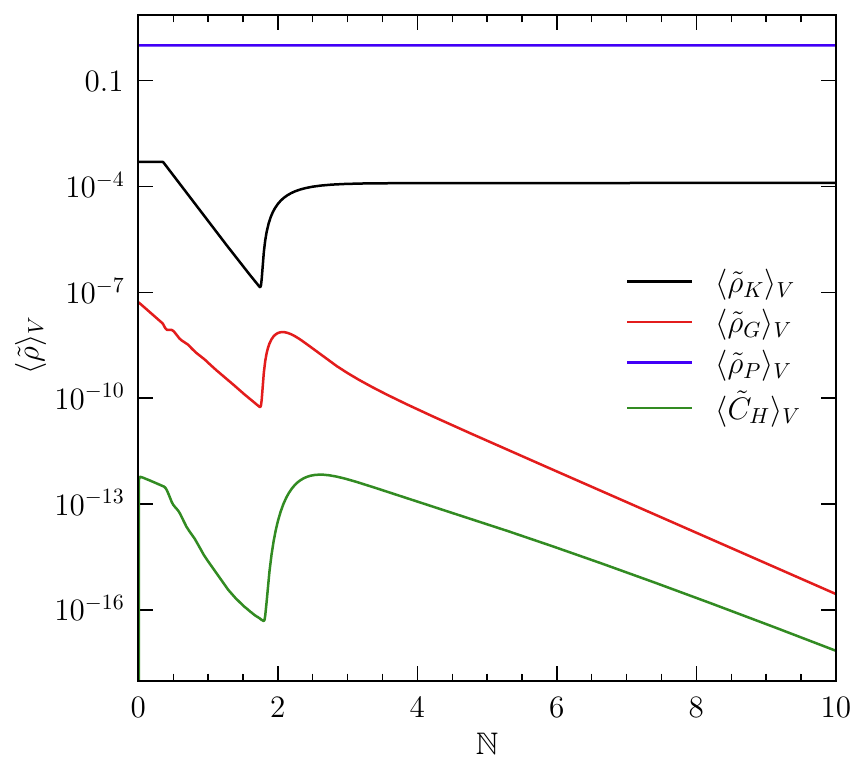}
    \end{minipage}
    \hfill
    \begin{minipage}{0.49\textwidth}
        \centering
        \includegraphics[width=\linewidth,height=0.85\linewidth,keepaspectratio]{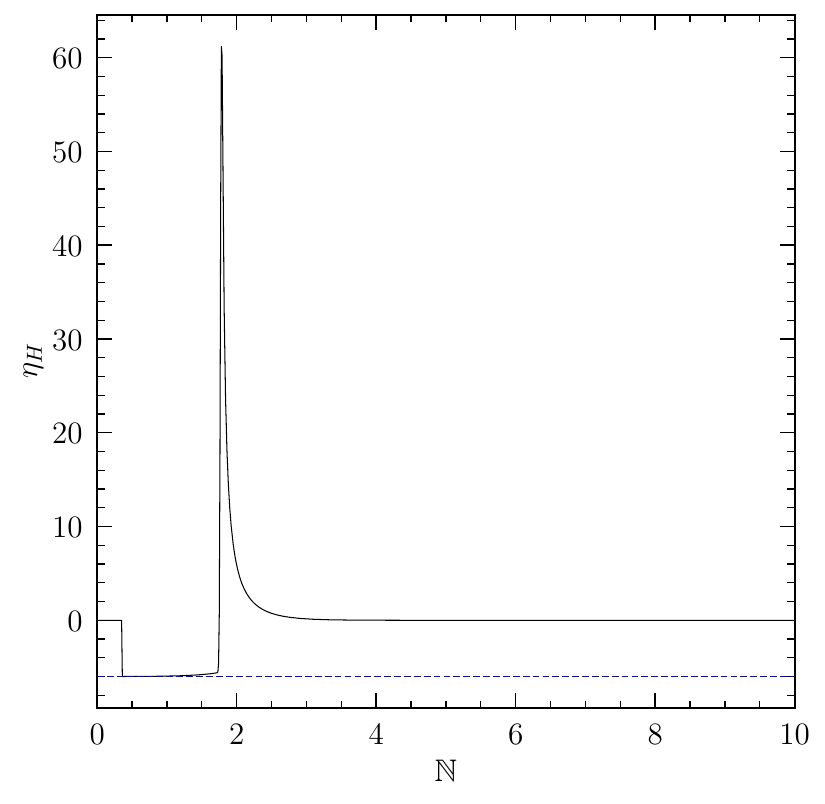}
    \end{minipage}
    \caption{
    We plot the evolution of the averaged energy components normalised by the total energy density and the effective second slow-roll parameters for the Starobinsky's linear model. The dashed line in the right panel indicates the \ac{USR} limit, $\eta_H=-6$.
    }
    \label{fig:figBGStaro3}
\end{figure}

The background evolution represented by Fig.~\ref{fig:figBGStaro3} is straightforward to interpret.  
Well before the break ($\phi \gg \phi_1$), the inflaton slow-rolls on the branch with slope $v_1$, yielding approximately constant slow-roll parameters set by $v_1$ and $V_0$. 
As the field approaches $\phi_1$, the effective slope drops to $v_1/\Lambda_1$, the inflaton temporarily overshoots the new attractor, and the system experiences a brief violation of the slow-roll conditions.  Depending on the choice of $\Lambda_1$ and the initial velocity, this can generate a transient dip in $\epsilon_H$ and a burst of enhanced curvature fluctuations around the scale $k_\star$ that exits the horizon when $\phi \approx \phi_1$. 
After the transition, the field settles back to the final slow-roll regime where the value of the slow-roll parameters is determined by the relative value of $\Lambda_1$ and $\Lambda_2$.

We implement this potential exactly as in Eq.~(\ref{eq:star3}), with the same piecewise form in simulations.
To verify that the discontinuity in the slopes 
does not introduce any numerical artifacts, we have also used the smoothed version of the potential~\cite{Martin:2011sn}.
This ensures that any features in the power spectrum or non-Gaussian statistics arise purely from the nonlinear field dynamics and the metric response, rather than from inconsistencies between the background and perturbation potentials. 
The numerical values of the parameters are chosen to produce the known results in the literature~\cite{Mizuguchi:2024kbl}:
$V_0 = 3\Mpl^2H_0^2$ with $H_0 = 10^{-5}\Mpl$.
The slope of the potential in the first slow-roll regime is chosen as $v_1 = \sqrt{9H_0^2/4\pi^2\Delta^2_\calR}$  where we set $\Delta^2_\calR= 8.5\times{}10^{-10}$.
The rest of the parameters are chosen to be $\phi_1 = 0.0$, $\phi_2 = -0.018$, $v_2 = v_1/850$ and $v_3 = v_1/2$, with the initial field value $\phi_{0} = 0.0193$, and the initial field derivative is chosen from the slow-roll attractor solution.

\begin{figure} 
    \begin{center}
    \begin{minipage}{0.325\textwidth}
        \centering
        \includegraphics[width=\linewidth,height=0.95\linewidth,keepaspectratio]{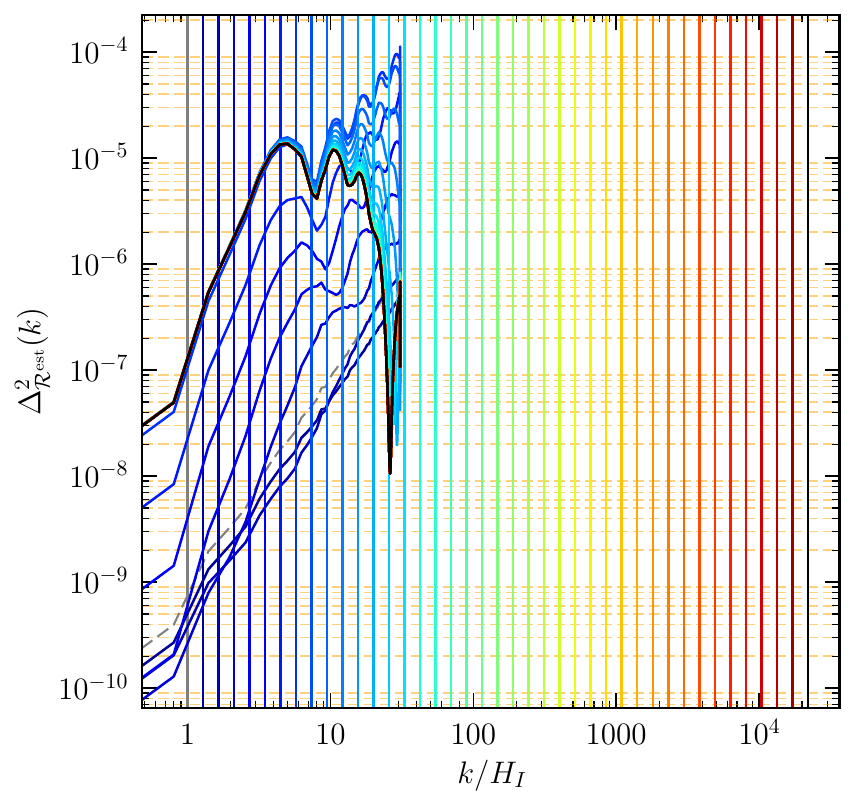} 
    \end{minipage}
    \begin{minipage}{0.325\textwidth}
        \centering
        \includegraphics[width=\linewidth,height=0.95\linewidth,keepaspectratio]{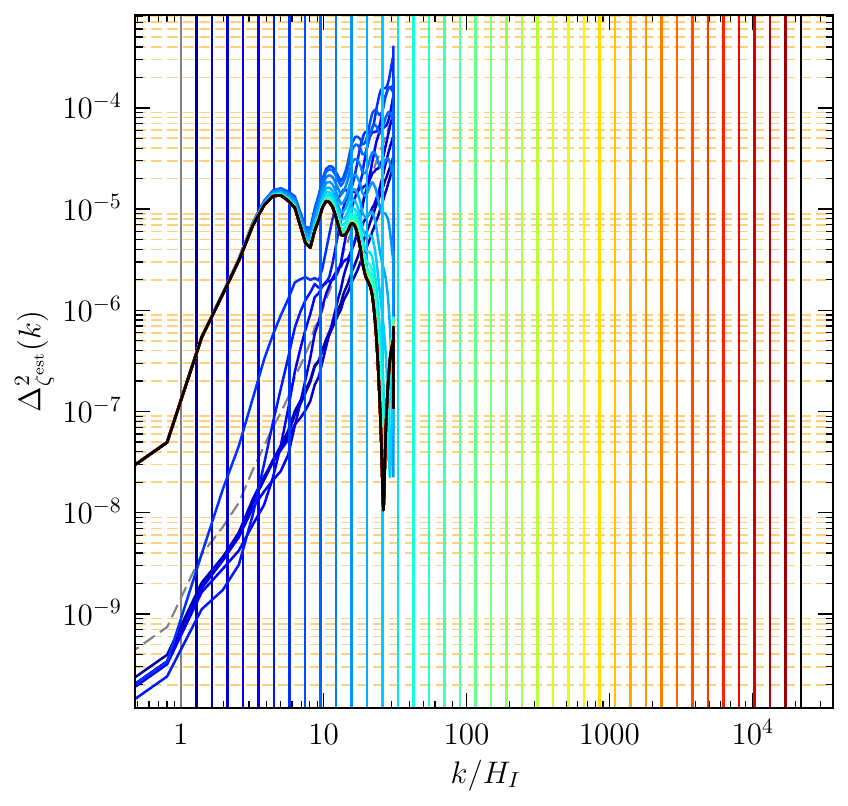} 
    \end{minipage}
        \begin{minipage}{0.325\textwidth}
        \centering
        \includegraphics[width=\linewidth,height=0.95\linewidth,keepaspectratio]{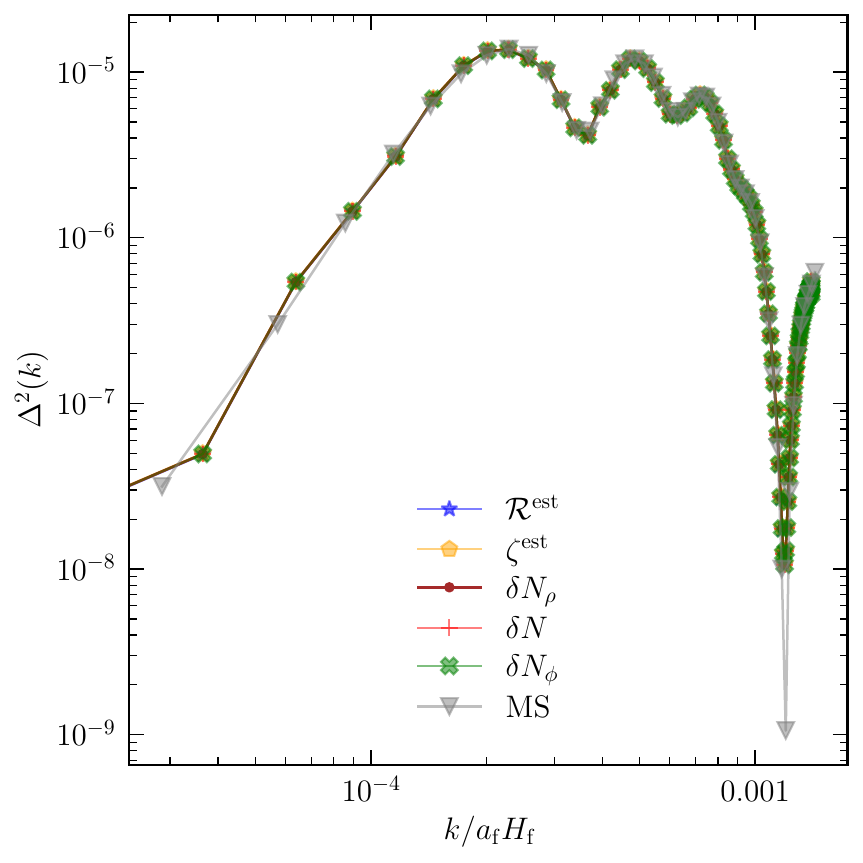} 
    \end{minipage}
    \end{center}
    \caption{
    We plot the time evolution of the power spectra for the comoving curvature perturbation $\mathcal{R}^{\mathrm{est}}$ (left panel) and the curvature perturbation on uniform-density slices $\zeta^{\mathrm{est}}$ (middle panel) during the initial SR, intermediate USR phase to the final SR phase, at intervals of $0.25$ \efold.
    The inflaton crosses the branch points $\phi_1$ at around $\mathbb{N} = 0.5$ and $\phi_2$ around $\mathbb{N} = 2$, which determines the duration of the USR phase during which the spectra of these two estimators don't converge.
    The colored vertical lines indicate the instantaneous comoving Hubble scale ($k = aH$) corresponding to the spectrum of the same colour. 
    For any given snapshot, modes to the left of the respective vertical line are super-Hubble, while those to the right are sub-Hubble. 
    At the onset of the simulation, the majority of the resolved spectrum is sub-Hubble, which is required to properly impose the adiabatic vacuum initial conditions for the field fluctuations. 
    By the end of the simulation (solid black line), the entire dynamical range of the lattice has crossed the horizon, leaving all modes deeply super-Hubble.
    The rightmost figure shows the final spectrum, along with spectra obtained using the $\delta{N}$ formalism. 
    }
    \label{fig:SpectraStaro3}
\end{figure} 

In Fig.~\ref{fig:SpectraStaro3}, we show the time evolution of the power spectra for $\mathcal{R}^{\mathrm{est}}$ and $\zeta^{\mathrm{est}}$ in the USR model. 
Each coloured curve corresponds to a snapshot separated by $\Delta\mathbb{N} = 0.25$ \efold, and the vertical line of the same colour marks the comoving Hubble scale $k = \bar{a}H$ at that time.  
Modes to the left of a given vertical line are super-Hubble at that snapshot, while those to the right remain sub-Hubble. 
During the initial slow-roll phase, $\mathcal{R}^{\mathrm{est}}$
and $\zeta^{\mathrm{est}}$ becomes constant after the horizon crossing, and they coincide with each other, as is well known from the perturbation theory.
However, the USR phase is governed by non-attractor background dynamics where the inflaton velocity evolves as $\dot{\phi} \propto 1/\bar{a}^{3}$. 
This violent deceleration prevents the super-horizon modes from freezing, leading to rapid growth in their amplitudes instead. 
Since the difference between $\mathcal{R}^{\mathrm{est}}$ and $\zeta^{\mathrm{est}}$ roughly scales as 
$\zeta^{\mathrm{est}} - \mathcal{R}^{\mathrm{est}} \propto \dot{\mathcal{R}}^{\mathrm{est}}/3H$, they deviate during this period.
After the field exits the plateau and re-enters the second slow-roll regime, the time variation of $\mathcal{R}^{\mathrm{est}}$ and $\zeta^{\mathrm{est}}$ starts to be suppressed again, and the two spectra gradually converge on super-Hubble scales.  
The evolving spectra in Fig.~\ref{fig:SpectraStaro3} visibly capture this non-trivial kinematic separation between the two curvature perturbation estimators during the USR phase.

The same interval in which the two curvature perturbation estimators diverge is also the regime in which the perturbations are expected to depart most strongly from Gaussianity: the breakdown of the slow-roll attractor during USR permits super-Hubble growth of the curvature perturbation and enhances nonlinear mode coupling. 
In the following, we therefore move beyond the power spectrum and study the full one-point \ac{PDF} of $\mathcal{R}^{\mathrm{est}}$ and $\zeta^{\mathrm{est}}$
on the lattice, extracting effective local-type non-Gaussianity parameters and tracking how the USR dynamics imprint a characteristic, strongly non-Gaussian signature on the curvature perturbation.


\begin{figure} 
    \begin{minipage}
    {0.47\linewidth}
    \begin{center}
        \includegraphics[width=1\linewidth]{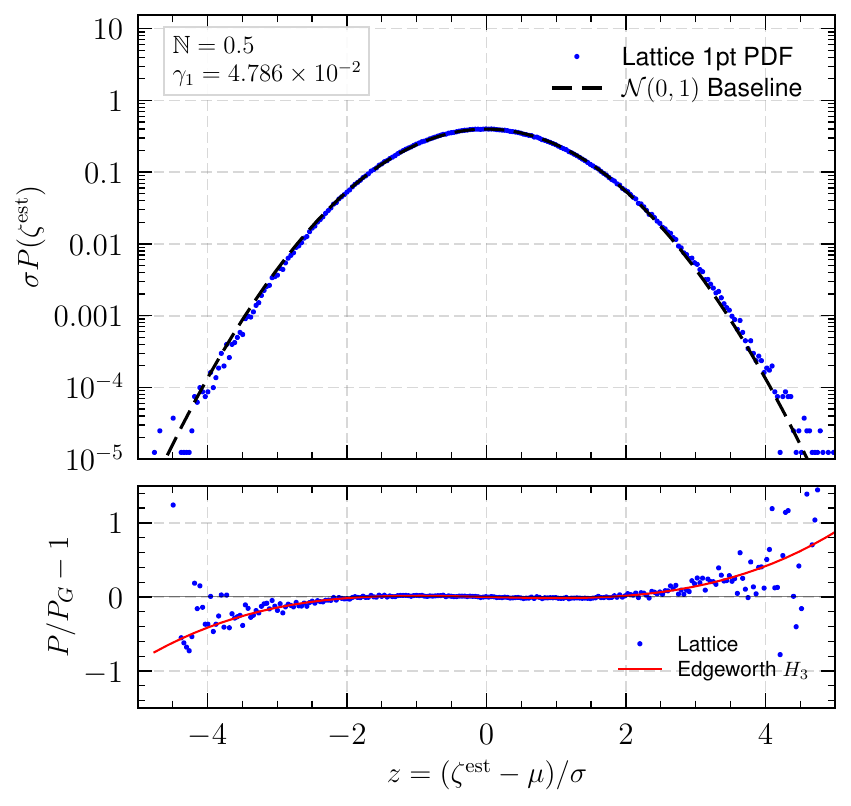} 
        \label{fig:pdf1}
    \end{center} 
    \end{minipage}
    \hfill
    \vspace{0.2 cm}
    \begin{minipage}
    {0.47\linewidth}
    \begin{center}
        \includegraphics[width=1\linewidth]{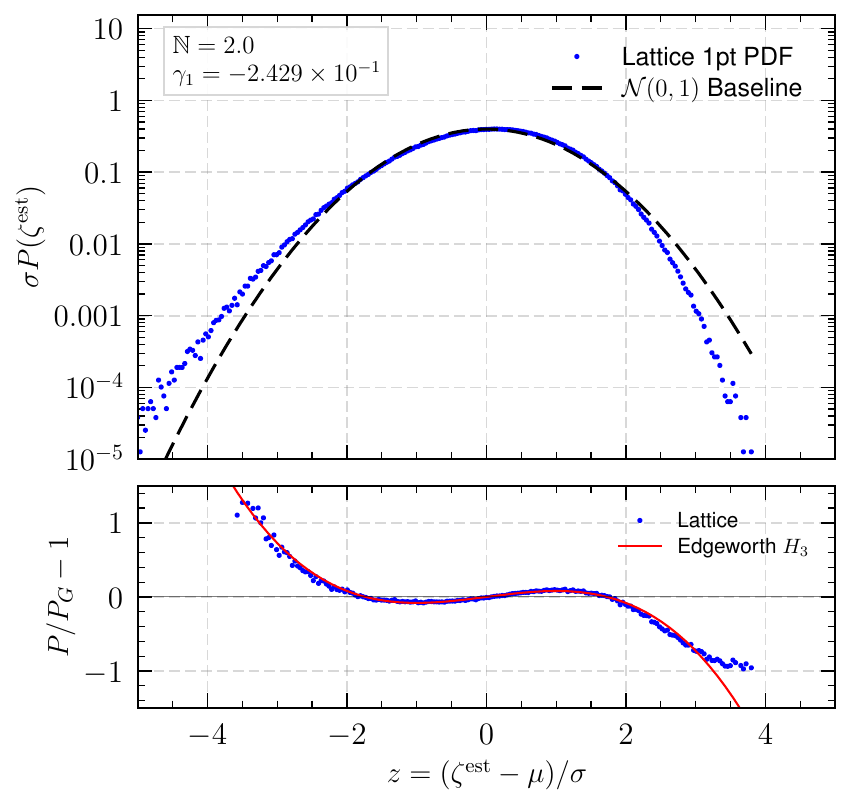} 
        \label{fig:pdf2}
    \end{center}
    \end{minipage}
    \vfill
    \vspace{0.2 cm}
    \begin{minipage}
    {0.47\linewidth}
    \begin{center}
        \includegraphics[width=1\linewidth]{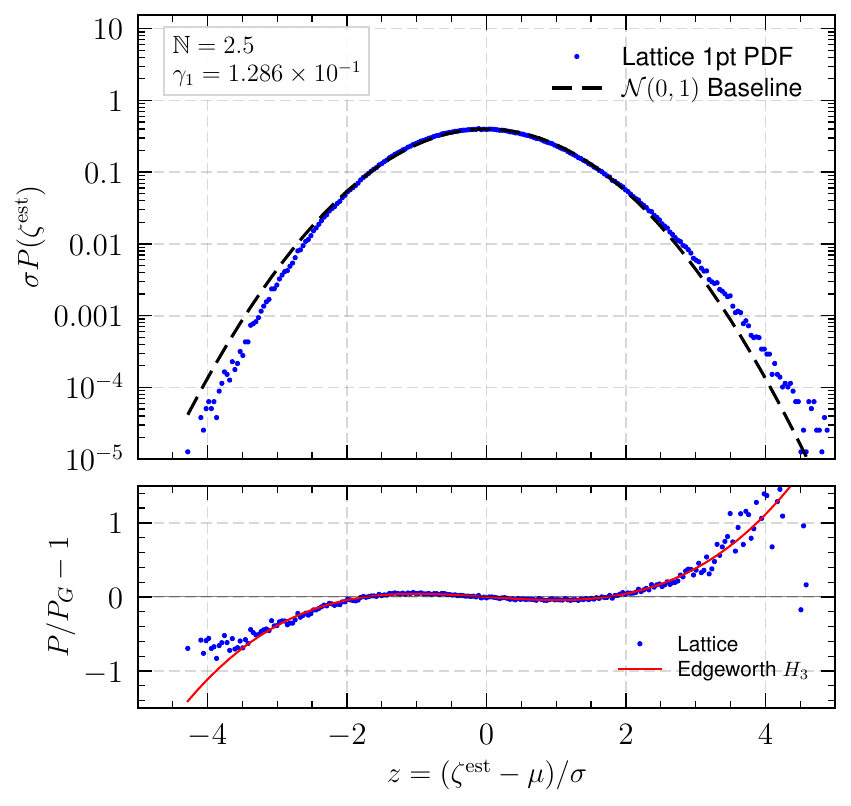} 
        \label{fig:pdf3}
    \end{center}
    \end{minipage} 
    \hfill
    \begin{minipage}
    {0.47\linewidth}
    \begin{center}
        \includegraphics[width=1\linewidth]{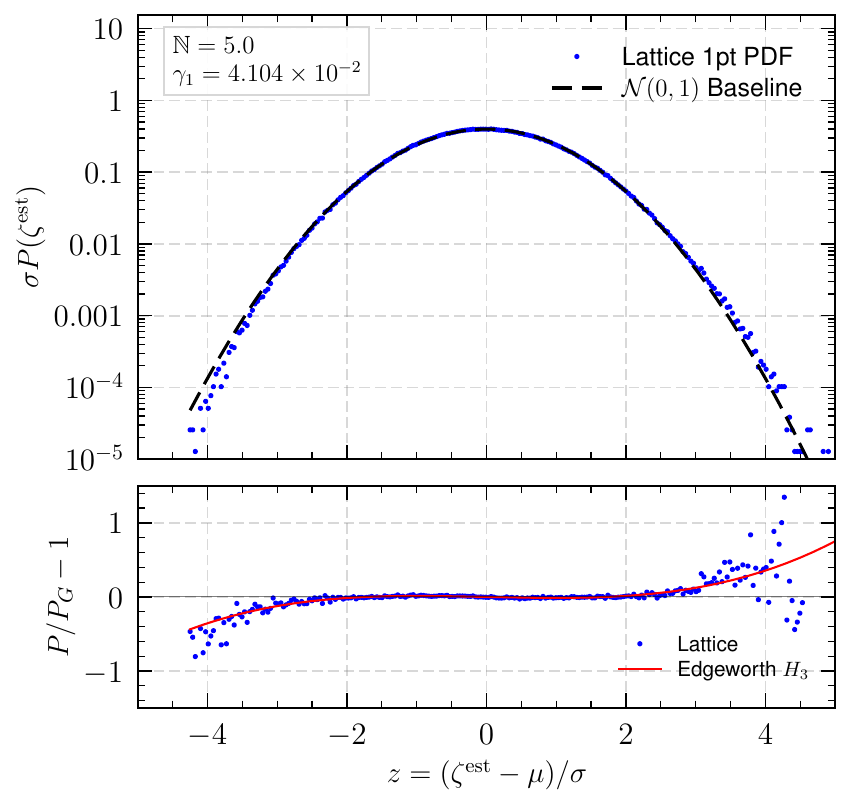} 
        \label{fig:pdf4}
    \end{center}
    \end{minipage}
    \caption{
    We plot the evolution of the standardised 1-point PDF of the curvature perturbation $\zeta^{\mathrm{est}}$.
    \emph{Top}: the standardised probability density $\sigma P(\zeta^{\mathrm{est}})$ plotted against the normalised amplitude $z = (\zeta^{\mathrm{est}}-\mu)/\sigma$ at four different instances. 
    We have also added the standard normal Gaussian baseline $\mathcal{N}(0,1)$ (dashed black) for comparison. 
    \emph{Bottom}: the fractional deviation from Gaussianity, $P/P_G - 1$. 
    The blue points denote the exact numerical lattice data, masked in the deep tails ($P < 10^{-6}$) to suppress finite-volume cosmic variance. 
    The red lines represent the theoretical 3rd-order Edgeworth expansion, dynamically scaled by the instantaneous lattice skewness $\gamma_1$. 
    The tight agreement in the core confirms the capture of the local non-Gaussian Hermite $H_3(z)$ signature, which permanently freezes into the grid upon entering the final SR phase.
}
\label{fig:pdf_4}
\end{figure}

To properly track the non-Gaussian evolution of the curvature perturbation $\zeta^{\mathrm{est}}$, as a prototype, across the \ac{USR} transition, we analyse its one-point \ac{PDF}, $P(\zeta^{\mathrm{est}})$, at several representative epochs: just before the USR plateau, at the end of the USR phase, shortly after re-entry into slow-roll, and deep in the final slow-roll regime (see Fig.~\ref{fig:pdf_4}). 
During USR, the variance, $\sigma^2 \equiv \langle (\zeta^{\mathrm{est}})^2 \rangle - \langle \zeta^{\mathrm{est}} \rangle^2$, grows rapidly. 
To disentangle this trivial variance amplification from the genuinely non-Gaussian evolution of the shape, we transform to standardised variables and plot the rescaled PDF $\sigma P(\zeta^{\mathrm{est}})$ as a function of the normalised $z$-score
\begin{equation}
    z \;\equiv\; \frac{\zeta^{\mathrm{est}} - \mu}{\sigma},
    \qquad
    \mu \equiv \langle\zeta^{\mathrm{est}}\rangle.
\end{equation}
In these variables, a purely Gaussian distribution of arbitrary variance collapses to the same fixed reference curve, $\mathcal{N}(0,1)$, so that any time evolution of the standardised PDF directly reflects departures from Gaussianity rather than simple growth of $\sigma$.
As shown in the upper panels of Fig.~\ref{fig:pdf_4}, the early-time distribution is nearly indistinguishable from the standard normal, while around the end of the USR plateau, pronounced asymmetric tails develop and persist into the subsequent evolution.

To quantify this departure from Gaussianity, the lower panel of Fig.~\ref{fig:pdf_4} shows the fractional deviation of the lattice PDF from the Gaussian baseline,
\begin{equation}
    \frac{\Delta P}{P_G}(z)
    \;\equiv\; \frac{P(\zeta^{\mathrm{est}})}{P_G(\zeta^{\mathrm{est}})} - 1,
\end{equation}
where $P_G$ is the standard normal. 
For a distribution characterised by \emph{mild} local-type non-Gaussianity, this deviation is well approximated by the Edgeworth expansion. 
Truncating at the leading non-Gaussian order, one obtains
\begin{equation}
    \frac{P(\zeta^{\mathrm{est}})}{P_G(\zeta^{\mathrm{est}})} - 1
    \;\simeq\; \frac{\gamma_1}{6}\,H_3(z)
    \;=\; \frac{\gamma_1}{6}\,(z^3 - 3z),
    \label{eq:edgeworth_H3}
\end{equation}
where $\gamma_1 \equiv \langle(\zeta^{\mathrm{est}}-\mu)^3\rangle/\sigma^3$ is the standardised skewness and $H_3(z)$ is the third probabilist’s Hermite polynomial. We overlay this theoretical $H_3$ “cubic wave” (black dotted line) directly on top of the numerical residuals $\Delta P/P_G$ from the lattice (blue curve). In the core region, $|z|\lesssim 3$, the agreement is excellent: the lattice curve tracks the predicted shape and crosses zero at the expected roots of $H_3(z)$, indicating that the USR dynamics have generated a positive skewness consistent with a positive local-type $f_{\rm NL}$ (to be quantified in the next subsection).
In the far tails ($|z|\gtrsim 3$), the lattice signal departs from the leading-order Edgeworth prediction, as higher moments and fully nonlinear effects become important. 
In this regime, we also mask the most extreme bins, where the PDF is dominated by a handful of rare cells and the finite simulation volume introduces substantial sampling (cosmic-variance) noise. 
After the field exits the plateau and the system settles into the final slow-roll attractor, both the core Hermite wave and the non-perturbative tails rapidly stabilise: the standardised PDF and its residual $\Delta P/P_G$ become time-independent within numerical accuracy. 
This freezing of the entire one-point distribution provides a direct, fully nonlinear confirmation of the super-horizon conservation of the curvature perturbation PDF in the final slow-roll phase.

To further quantify the non-Gaussianity for the transitions from SR to USR and eventually to the SR phase considered here, we calculate the associated $f_{\mathrm{NL}}$ parameter as shown in Eq.~\eqref{eq:fNL_from_S3}.
We find that the effective $f_{\rm NL}^{\rm(1pt)}$ exhibits the following characteristic behaviour as shown in Fig.~\ref{fig:figStaro3fNL}. During the initial slow-roll phase, the skewness is numerically consistent with zero and $f_{\rm NL}^{\rm(1pt)}\approx 0$, in agreement with the standard single-field slow-roll result $f_{\rm NL}^{\rm local}\sim \mathcal{O}(1-n_s)$. As the background enters the non-attractor \ac{USR} plateau, the one-point PDF of $\zeta^{\mathrm{est}}$ develops a pronounced positive skewness and the estimator~\eqref{eq:fNL_from_S3} asymptotes to
\begin{equation}
  f_{\rm NL}^{\rm(1pt)} \;\to\; \frac{5}{2},
\end{equation}
for scales that exit the horizon during the USR phase. 
Once the trajectory exits USR and returns to a slow-roll regime, the growth of $\zeta^{\mathrm{est}}$ is halted, the skewness ceases to evolve, consequently $f_{\rm NL}^{\rm(1pt)}$ freezes close to $5/2$ for modes exiting during USR and remains constant thereafter.
This plateau at $f_{\rm NL}\simeq 5/2$ for modes exiting during the USR phase is in agreement with analytic treatments of single-field \ac{USR} inflation in comoving gauge, which predicts an effective local-type non-Gaussianity $f_{\rm NL}=5/2$ for such modes~\cite{Namjoo:2012aa,Martin:2012pe,Chen:2013aj,Cai:2018dkf,Pi:2022ysn}, and it is consistent with the soft-theorem analyses of shift-symmetric cosmologies~\cite{Finelli:2017fml}, which clarify how a non-decaying local bispectrum can arise in USR without contradicting single-field consistency relations. 
\begin{figure} 
    \begin{center}
    \begin{minipage}{0.48\textwidth}
        \centering
        \includegraphics[width=\linewidth,height=0.85\linewidth,keepaspectratio]{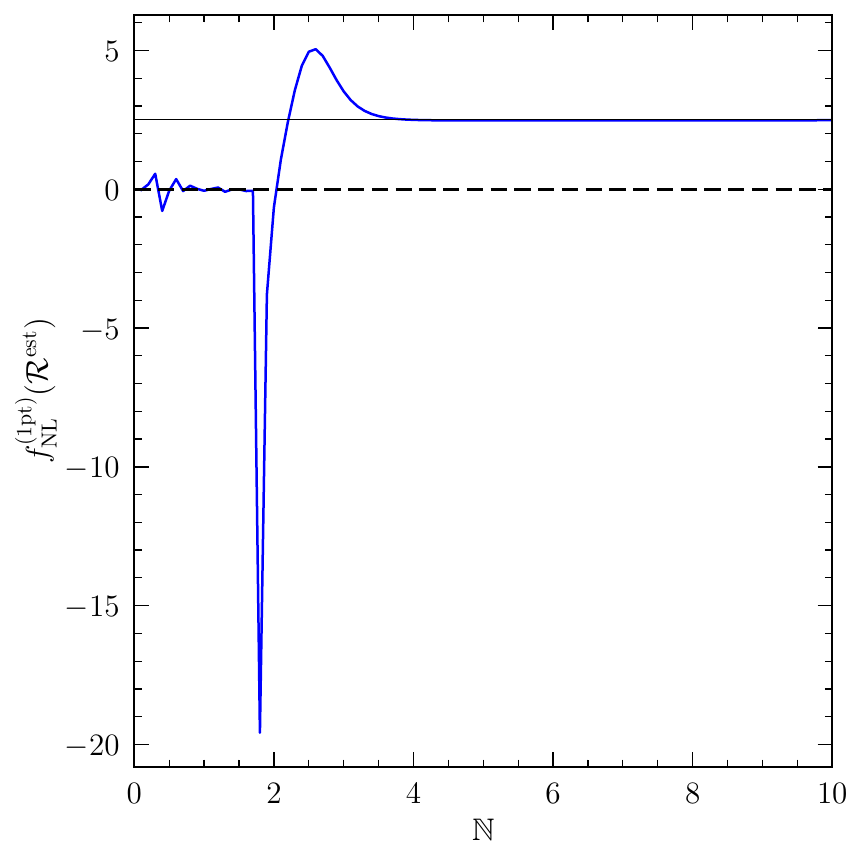} 
    \end{minipage}
    \begin{minipage}{0.48\textwidth}
        \centering
        \includegraphics[width=\linewidth,height=0.85\linewidth,keepaspectratio]{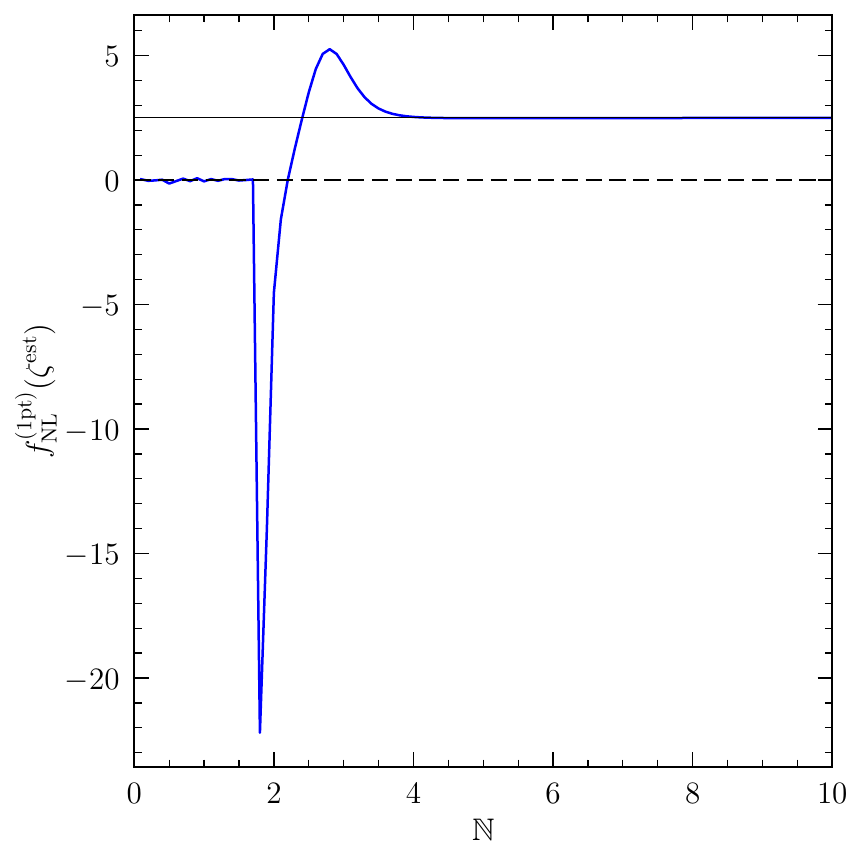} 
    \end{minipage}
    \end{center}
    \caption{
    The evolution of the effective one-point $f_{\mathrm{NL}}$ for the linear estimator for comoving curvature perturbation $\mathcal{R}^{\mathrm{est}}$ (left) and curvature perturbation on uniform density hypersurface $\zeta^{\mathrm{est}}$ (right).
    }
    \label{fig:figStaro3fNL}
\end{figure} 
In concrete realisations, $f_{\mathrm{NL}}$ can acquire a mild scale dependence for modes that exit the horizon close to the end of the USR phase, such that the value deviates from the pure plateau value of $5/2$~\cite{Passaglia:2018ixg,Ragavendra:2020sop,Ozsoy:2021pws,Motohashi:2023syh,Namjoo:2025hrr,Escriva:2025ftp}.
Our lattice measurement of $f_{\rm NL}^{\rm(1pt)}$ from the real-space $\zeta^{\mathrm{est}}$ distribution reproduces this behaviour quantitatively: the skewness-based estimator tracks $f_{\rm NL}\simeq 0$ in slow-roll, rises to a plateau near $5/2$ through the USR window, and then freezes once the system re-enters the attractor SR regime. 
Within our numerical uncertainties (finite volume, lattice resolution, and the use of a one-point moment estimator rather than a full bispectrum reconstruction), this is fully consistent with the analytic prediction from the $\delta{N}$ formalism, indicating that the modes around or below the horizon scales did not play an important role in this model.
While the small residual offset from $5/2$ may reflect a combination of mild scale dependence and discretisation effects, we do not attempt to interpret it as a precise measurement of running.

\subsection{\label{sec:eval_shear}Validity of the shear-free approximation during USR}

\begin{figure} 
    \centering
    \begin{minipage}{0.49\textwidth}
        \centering
        \includegraphics[width=\linewidth,height=0.85\linewidth,keepaspectratio]{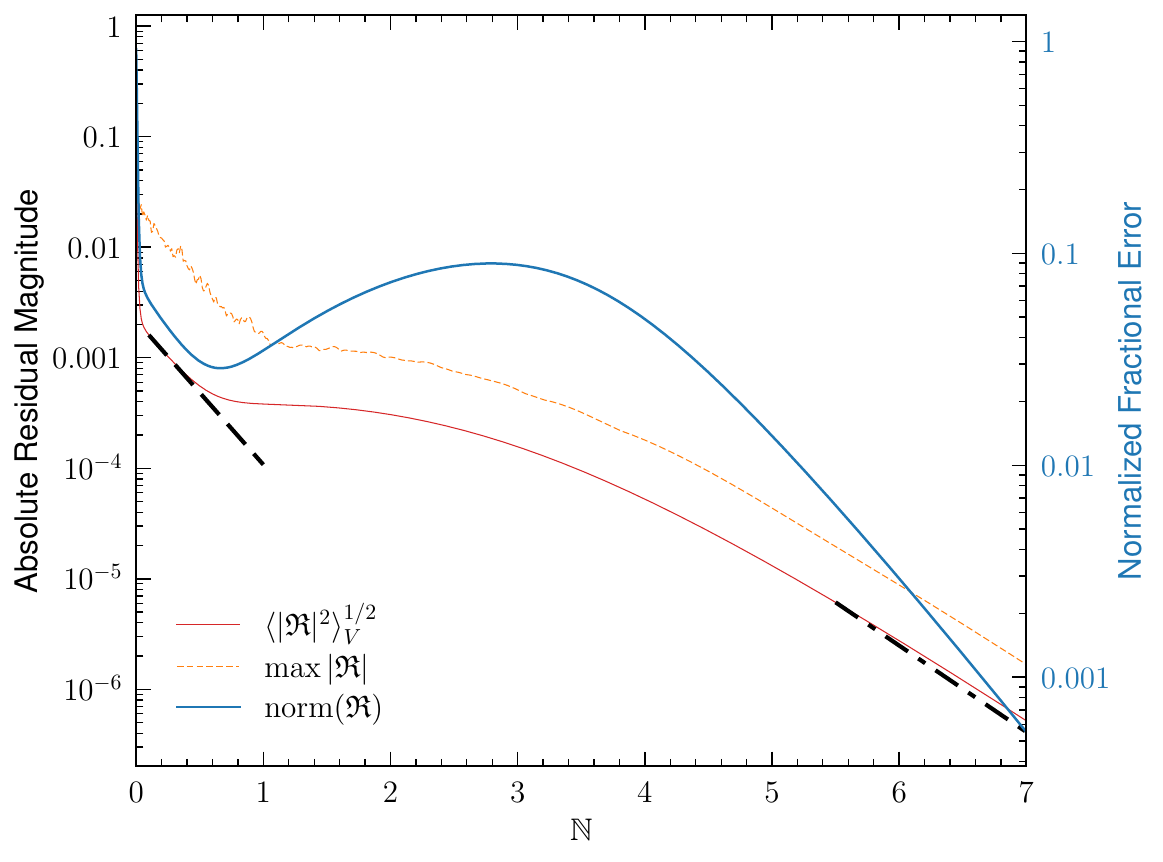}
    \end{minipage}
    \hfill
    \begin{minipage}{0.49\textwidth}
        \centering
    \includegraphics[width=\linewidth,height=0.85\linewidth,keepaspectratio]
    {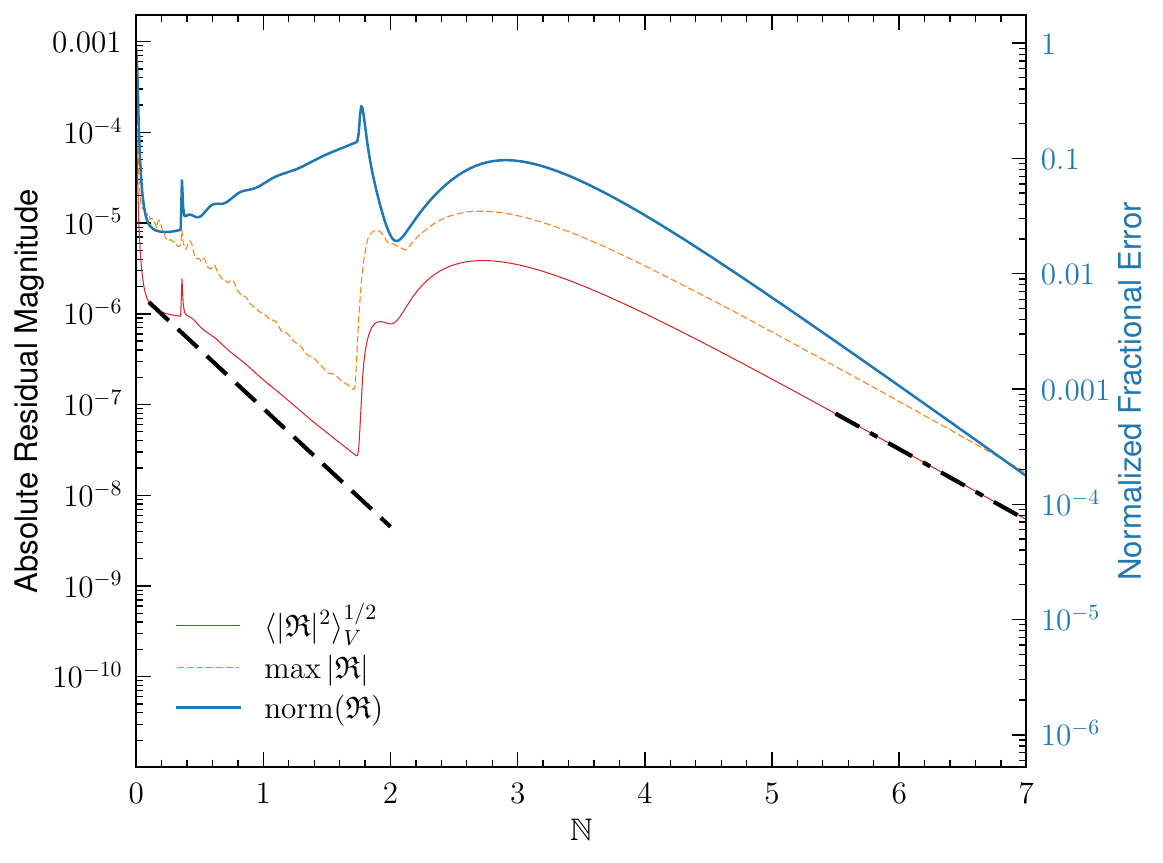}
    \end{minipage}
    \caption{
    We show the evolution of the momentum constraint violation and its scaling dynamics. 
    The left panel is for the slow-roll $m^2\phi^2$ model, and the right panel is for the Starobinsky's linear model.
    The left vertical axis corresponds to the absolute magnitudes of the residual vector $\mathfrak{R}$, showing both its volume-weighted \ac{rms} and maximum values across the grid. The right vertical axis maps the dimensionless normalised residual, $\mathrm{norm}(\mathfrak{R})$ (blue solid line). 
    The dashed lines depict the scaling behaviour as explained in the main text.
    }
    \label{fig:figMCStaro3}
\end{figure}
To complement our analysis, it is useful to track the evolution of the momentum-constraint residual as defined in Eq.~(\ref{eq:R_def}),  
whose norm provides a direct diagnostic of the error associated with the shear-free approximation (cf. Eq.~\eqref{eq:MC_full}).
At the early time in both models, the absolute \ac{rms} of the residual, 
$\langle|\mathfrak{R}|^2\rangle_V^{1/2}$ is found to decay as $e^{-3\mathbb{N}}$. For the USR model, this scaling remains intact throughout the USR phase.
This behavior is consistent with the result at the leading order of the gradient expansion~(see Sec.~\ref{sec:dec_shear}). 
This later crosses over to a slower $a^{-1.8}$ scaling once the modes of interest are well outside the Hubble radius, signaling that the leading contribution has become subdominant and that the next-order gradient terms are taking over. 
Including the shear degree of freedom would be expected to reduce the residual associated with the shear-free truncation. 
However, whether it fully restores the leading-order scaling once that contribution has redshifted away requires a dedicated analysis.

If we now turn to $\mathrm{norm}(\mathfrak R)$ (solid blue curves, plotted against the right-hand axis), the interpretation is slightly different. 
As defined in Eq.~\eqref{eq:normR}, this dimensionless quantity is constructed from the rms amplitudes of $\partial_i H (\equiv L_i)$ and $(2\Mpl^2)^{-1}\dot{\phi}\,\partial_i\phi (\equiv - R_i)$, and therefore measures the residual relative to the two dominant terms that are supposed to cancel in the momentum constraint. 
In the pure slow-roll model, the normalised residual initially decreases, then rises during the crossover away from the leading $e^{-3\mathbb{N}}$ regime, and finally decreases again once the late-time scaling is established. 
In the USR model, by contrast, $\mathrm{norm}(\mathfrak R)$ continues to grow throughout the USR phase even though the absolute residual is still decreasing. 
This is because the denominator, built from the rms amplitudes of $L_i$ and $R_i$, is suppressed more rapidly during USR, primarily due to the decay $\dot\phi \propto e^{-3\mathbb{N}}$. 
The resulting mismatch in scaling produces the transient $\mathcal O(0.5)$ enhancement of the normalised residual. 
Once the system exits USR and returns to the final slow-roll attractor, the normalised residual follows the same qualitative pattern as in the pure slow-roll model.

\section{\label{sec:conc}Conclusion and outlook}

In this work, we developed and implemented a lightweight but relativistically inhomogeneous framework for lattice simulations of inflation.   
Instead of evolving the full metric degrees of freedom as in BSSN-type formalism, we have adopted an inhomogeneous FLRW ansatz with a local scale factor $a(\mathbf{x},t) = \bar{a}(t)\,e^{\psi(\mathbf{x},t)}$ and a canonical scalar inflaton.
This setup retains the nonlinear scalar dynamics, together with the leading local gravitational response --- namely, the spatially varying Hubble rate, the curvature contribution to the local Friedmann constraint, and proper-volume weighting --- within the weakly nonlinear regime tested, while remaining much cheaper than full $3+1$ relativity on large lattices.  
Within this framework, we evolved the inhomogeneous scalar field, its conjugate momentum, and the local expansion field on a three-dimensional lattice, and used these solutions to construct curvature perturbations, $\delta N$ observables, and one-point non-Gaussian statistics.

We first validated the formalism in a simple slow-roll model with a quadratic potential, where genuine relativistic corrections are expected to remain small, and analytic estimates are confirmed with numerical results~\cite{Giblin:2019nuv,Caravano:2021pgc}.  
We further verified that different estimators of the curvature perturbation --- including $\zeta^{\mathrm{est}}$, $\mathcal{R}^{\mathrm{est}}$, and the fully nonlinear $\delta{N}$ construction on constant-density (or constant-field) slices --- agree very well on super-Hubble scales, confirming that the implementation reproduces the expected attractor behaviour. 
We then applied the same machinery to Starobinsky's linear-potential model as a representative of a model beyond slow-roll dynamics.
For this model featuring a USR plateau between two SR regimes, the lattice captures the temporary separation between different curvature perturbation estimators, the growth and subsequent freezing of non-Gaussianity measures, and the time variation of the momentum-constraint residual. In particular, the simulations show that the violation of the momentum constraints remains small, while the normalised residual is temporarily enhanced during the deepest part of the USR phase, signaling a transient weakening of the shear-free approximation precisely when the inflaton velocity becomes very small.
At later times, once the system re-enters the final slow-roll attractor, the curvature estimators converge again, and the full one-point PDF of the curvature perturbation freezes to a mildly non-Gaussian form consistent with an effective local-type $f_{\mathrm{NL}}\to 5/2$.

The scenarios studied here were deliberately conservative: we focused on models in which relativistic effects are expected to remain modest, so that curvature and constraint diagnostics primarily serve as internal consistency checks and as a first validation of the method. 
The natural next step is to push the same framework into regimes where local gravitational effects are expected to become genuinely important, including models with very large small-scale power, extended USR phases, or sharp features engineered to enhance the non-Gaussian tails relevant for primordial-black-hole formation. 
It will also be important to extend the present setup to multi-field systems and to spectator sectors with nontrivial sound speeds, where isocurvature modes and non-adiabatic pressure can continue to source curvature perturbations after horizon exit.
Another important extension is to include gauge fields in the setup, for instance, to study axion-gauge couplings, where nonlinear dynamics can generate large anisotropic stresses, chiral field amplification, and the growth of sub-Hubble structure. 
Finally, systematic benchmark comparisons with fully relativistic simulations will be needed to map out the precise range of validity of the locally FLRW, shear-free approximation, and to determine when this lightweight GR framework remains reliable and when the full machinery of numerical relativity becomes unavoidable.

\acknowledgments
P.~S. is grateful to John T. Giblin Jr. for insightful and very helpful discussions during iTHEMS Cosmology Forum $\mathrm{n}^{\circ}3$ - (P)reheating the primordial Universe at RiKEN. Y.~U. would like to thank Jaume Garriga and Takahiro Tanaka for valuable discussions. All of the authors are supported by Grant-in-Aid for Scientific Research under Contract No.~JP26H00402(26H00402).
P.~S. and Y.~U. are supported by Grant-in-Aid for Scientific Research (B) under Contract No.~23K25873 (23H01177) and JST FOREST Program under Contract No. JPMJFR222Y. Y.~T. is supported by JSPS KAKENHI Grant
No.~JP24K07047. Y.~U. is also supported by Grant-in-Aid for Scientific Research under Contract No. JP21KK0050.
Numerical computation in this work was carried out at the Yukawa Institute Computer Facility.


\appendix

\section{\label{app:backreaction}Continuity equation and backreaction}
A well-known subtlety in averaging in general relativity is that the volume average of the local energy density does not, in general, satisfy a closed FLRW continuity equation. 
This issue is rigorously treated in the scalar averaging framework developed by Buchert for inhomogeneous cosmologies~\cite{Buchert:1999er,Buchert:2001sa,Buchert:2019mvq}.
For foundational treatments of cosmological backreaction and the extensive debate regarding its dynamical relevance to late-time cosmic acceleration, see, e.g.,~\cite{Ellis:1984bqf,Larena:2008be,Gasperini:2009wp,Wiltshire:2009db,Gasperini:2009mu,Baumann:2010tm,Green:2010qy,Green:2014aga,Buchert:2015iva}, and for comprehensive reviews, see~\cite{Rasanen:2011ki,Wiltshire:2011vy,Clarkson:2011zq,Debono:2016vkp,Fleury:2025ykl}. 
Furthermore, for classic treatments of how these effects manifest in the scalar-field dynamics during inflation, see~\cite{Abramo:1997hu,Geshnizjani:2002wp}.
In that formalism, spatial averages obey effective Friedmann equations sourced by a `backreaction' fluid, $\mathcal{Q}_\mathcal{D}$, which encodes the variance of the local expansion rate and shear.
Our lattice implementation adopts this framework precisely. The volume weights $w_i \propto \sqrt{\gamma_i}$ introduced in Eq.~\eqref{eq:latt_avg} provide the natural Riemann-sum version of the Buchert average on a grid.
Consequently, the time evolution of averaged quantities in our simulation is subject to the same backreaction effects.

Let $X(\mathbf{x},t)$ be any scalar on a constant-time hypersurface, and $\langle X\rangle_V$ the physical-volume average defined in Eq.~\eqref{eq:volavg_here}. Differentiating with respect to $t$ and using $\sqrt{\gamma} = a^3(\mathbf{x},t)$, one finds the general identity
\begin{equation}
  \frac{\dd}{\dd t}\langle X\rangle_V
  = \big\langle \dot X \big\rangle_V
    + 3\Big(\big\langle H X \big\rangle_V
            - \langle H\rangle_V \langle X\rangle_V\Big),
  \label{eq:dt_avg_general}
\end{equation}
where $H(\mathbf{x},t) \equiv \dot a(\mathbf{x},t)/a(\mathbf{x},t)$ is the local expansion rate. 
The second term on the right-hand side is proportional to the volume-weighted covariance between $H$ and $X$. 
It explicitly states that taking a time derivative and a volume average are, in general, non-commuting operations.
Now specialising to the local energy density $\rho(\mathbf{x},t)$ and pressure $p(\mathbf{x},t)$ of the scalar field, which obey the local continuity equation
\begin{equation}
  \dot{\rho}(\mathbf{x},t) + 3H(\mathbf{x},t)\bigl[\rho(\mathbf{x},t) + p(\mathbf{x},t)\bigr] = 0,
\end{equation}
and inserting $X=\rho$ into Eq.~\eqref{eq:dt_avg_general}, we obtain
\begin{equation}
  \frac{\dd}{\dd t}\langle\rho\rangle_V
  = -3\big\langle H(\rho+p)\big\rangle_V
    + 3\,\mathrm{Cov}_V\bigl(H,\rho\bigr),
  \label{eq:backreaction}
\end{equation}
with
\begin{equation}
  \mathrm{Cov}_V(H,\rho)
  \;\equiv\;
  \big\langle H \rho\big\rangle_V
  - \langle H\rangle_V\,\langle\rho\rangle_V.
\end{equation}
The first term on the right-hand side is what one would obtain by naively averaging a homogeneous continuity equation, while the covariance term encodes the effect of correlations between local expansion and density inhomogeneities. In the Buchert language, such correlation terms, together with the variance of $H$ and the shear, contribute to the effective backreaction source $\mathcal{Q}_\mathcal{D}$ that modifies the averaged Friedmann equations.

In this work, we do not enforce that the instantaneous volume average $\langle\rho\rangle_V$ obey a closed FLRW continuity equation of the form $\dot\rho + 3\bar H(\rho+p)=0$ with $\bar H = \langle H\rangle_V$. 
Strictly speaking, one could define an integrated effective background energy density $\rho_{\rm bg}$ that perfectly satisfies the averaged local conservation law,
\begin{equation}
  \dot{\rho}_{\rm bg}(t)
  \;\equiv\;
  -3\,\big\langle H(\rho+p)\big\rangle_V.
  \label{eq:rho_bg_def}
\end{equation}
Equation~\eqref{eq:rho_bg_def} is the lattice analogue of the averaged continuity equation in Buchert's formalism, with the backreaction contributions contained in the discrepancy between the theoretical $\rho_{\rm bg}(t)$ and the naive instantaneous average $\langle\rho\rangle_V(t)$.

However, for the simulations presented in this paper, the scalar field gradient and curvature contributions to the total energy budget remain small, and the residuals of the Hamiltonian and momentum constraints are numerically negligible. 
In this regime, the covariance term in Eq.~\eqref{eq:backreaction} is found to be much smaller than the leading term, ensuring that the theoretical $\rho_{\rm bg}(t)$ closely tracks the instantaneous volume average $\langle\rho\rangle_V(t)$. 
Because the backreaction is negligible, we are justified in using the direct instantaneous volume averages, $\bar\rho \equiv \langle\rho\rangle_V$ and $\bar p \equiv \langle p\rangle_V$, to compute the effective Hubble slow-roll parameters and background quantities quoted in the main text. 
This allows us to efficiently characterise the evolution of the inhomogeneous domain using variables computed directly on the lattice at each time slice.

\section{\label{app:lattice_details}Lattice equations, initialisation, and averaging}

In this appendix, we will describe the equations solved in the lattice, the rescaling program, and their implementation details. 
For this purpose, we developed a dedicated module within our broader simulation suite, \textsc{LattE} (\textbf{Latt}ice \textbf{E}volver), a high-performance \texttt{C++} framework designed for real-space simulations of nonlinear early-Universe dynamics. 
The broader \textsc{LattE} architecture also includes existing modules for gauge fields and gravitational-wave production, the applications of which will be reported separately.
The implementation used here, \textsc{LattEfold} (\textbf{Latt}ice \textbf{E}volver in E-folds), evolves the Einstein--scalar system on a three-dimensional periodic lattice using the background \efold variable as the time coordinate.
To ensure computational efficiency across large grid volumes, the core evolution loops are parallelised via OpenMP for shared-memory execution, and the suite relies on the highly optimised FFTW library to evaluate all required discrete Fourier transforms~\cite{frigo2005design}.

\subsection{\label{app:vars}Dynamical variables and units}

In this work, we consider a single field inflation described by a potential $V(\phi)$ and canonical kinetic term evolving on an expanding background with a scalar metric perturbation $\psi(\mathbf{x},t)$ that encodes local expansion. 
Throughout, we assume a cubic comoving box of side length $L$, discretised into $\Ng^3$ points with lattice spacing $\Delta x = L/\Ng$, and periodic boundary conditions. 
We have used the following variable rescalings:
\begin{align}
    \phi_{\mathrm{pr}} \equiv \frac{\phi}{\Mpl};\quad\pi_{\phi,\mathrm{pr}}\equiv\frac{\pi_\phi}{\Mpl};\quad\vec{x}_{\mathrm{pr}}\equiv B\vec{x};\quad H_{\mathrm{pr}}\equiv\frac{H}{B},
\end{align}
where $\pi_\phi = \dot{\phi} = \bar{H}\!\dv*{\phi}{\mathbb{N}}$. 
The mass scale $B$ is chosen from the characteristic mass scale of the model; for instance, for $V(\phi) = m^2\phi^2/2$, it is simply the inflaton mass $m$. 
The corresponding dimensionless potential and energy densities are:
\begin{equation}
    V_{\mathrm{pr}} = \frac{V}{\Mpl^2B^2}\quad \rho_{\mathrm{pr},i} = \frac{\rho_i}{\Mpl^2B^2}
\end{equation}
Now, for brevity, we will omit the subscript `pr' from all the variables in the subsequent discussion. However, from now on, we are considering the program variables.
Most of the lattice details and discretisation thus follow the standard lattice implementation~\cite{Felder:2007les,Figueroa:2021yhd,Baeza-Ballesteros:2025tme}.
Below, we mention some of the quantities that are different from the usual rigid-FLRW Universe.
Further, we also kept an option to simulate in the rigid-FLRW case for benchmarking against standard lattice codes.

\subsubsection{Local energy, Hubble rate, and curvature term}
For a given configuration $(\phi,\pi_\phi,\mathbb{N})$ we define the local program-unit energy density
\begin{equation}
\rho(x) \;=\;
\frac{1}{2}\,\pi_\phi^2(x)
\;+\; \frac{1}{2}\,e^{-2N_{\rm geom}(x)}\,|\nabla\phi(x)|^2
\;+\; V\!\left(\phi(x)\right),
\label{eq:rho_prog_full}
\end{equation}
where $N_{\mathrm{geom}}$ is the ``geometry'' used inside spatial operators $N_{\mathrm{geom}}(x) \equiv \mathbb{N} + \psi(x)$. 

The local Hubble rate is then defined from a generalised Hamiltonian constraint,
\begin{equation}
H^2(x) = \frac{\rho_{\mathrm{pr}}(x)}{3}
\;+\; C_H(x),
\label{eq:Hx2_def}
\end{equation}
where the second term is a “curvature” or gradient correction constructed from $N$,
\begin{equation}
C_H(x) = \frac{2}{3}\,e^{-2N_{\rm geom}(x)}
\left[ \nabla^2 \psi(x) + \frac12 |\nabla \psi(x)|^2 \right].
\label{eq:CH_def}
\end{equation}
When running for rigid-FLRW ($\psi = 0$), we got $C_H=0$. 
The background Hubble rate $\bar H$ is evolved separately via a Raychaudhuri equation (see below).

\subsection[Evolution in background \efolds]{\boldmath \label{app:latt_Eqns}Evolution in background \texorpdfstring{\efolds}{efolds}}

To represent a continuous field on the lattice, we assign its value to each site of a cubic grid with \(\Ng^3\) points. 
Thus, any scalar field \(X(\mathbf{x},\mathbb{N})\) in the continuum is replaced by the discrete set
\begin{equation}
  X(\mathbf{x},\mathbb{N})
  \;\longrightarrow\;
  X_{i,j,k}(\mathbb{N}),
  \qquad
  i,j,k \in \{1,\dots,N\},
\end{equation}
where \((i,j,k)\) labels the lattice site. 

We solve the equations with the background number of \efolds as the \emph{time}-variable
$$
\mathbb{N}(t) \equiv \ln \bar{a}(t) ,
$$
so that $\dv*{\mathbb{N}} = \bar{H}^{-1} \dv*{t}$, where $\bar{H}$ is the background Hubble rate defined only through the lattice averaged quantities. In the following, we use Primes to denote derivatives with respect to $\mathbb{N}$, e.g.
\begin{align}
    X' \equiv \dv{X}{\mathbb{N}} = \frac{1}{\bar H}\dv{X}{t}.
\end{align}
In addition to the usual evolution equations for the scalar field $\phi$ and its conjugate velocity $\pi_{\phi} \equiv \bar{H}\phi'$, the code also evolves a scalar metric variable $N_{\mathrm{geom}}$, which measures the local \efolds relative to the background and controls the local expansion rate $H$.
Thus, we take the basic dynamical variables to be
\begin{align}
\phi(\mathbf{x},\mathbb{N}),\quad \pi(\mathbf{x},\mathbb{N})\equiv\dot\phi(\mathbf{x},t(\mathbb{N})),\quad N_{\mathrm{geom}}(\mathbf{x},\mathbb{N}),\quad \bar H(\mathbb{N}).
\end{align}
We will suppress the explicit lattice-site labels $(i,j,k)$ and the explicit time dependence $(\mathbb{N})$ in the subsequent equations, unless they are not clear from the context.

\subsubsection{Field equations}

Using \eqref{eq:KGpFLRW}, the lattice evolution equations are
\begin{align}
\phi' &= \frac{\pi_{\phi}}{\bar H},
\label{eq:phi_prime}\\
\pi_{\phi}'  &= -3\,\frac{H_{\rm fric}}{\bar H}\,\pi_\phi + \frac{e^{-2N_{\rm geom}}}{\bar H} \left[\nabla^2\phi + \nabla \psi\cdot\nabla\phi\right] - \frac{1}{\bar H}\,V_{\,\phi},
\label{eq:pi_prime}
\end{align}
where $H_{\rm fric}$ is the `friction' Hubble rate that appears in the local \ac{KG} equation. 
In our implementation, we can choose either
\begin{align}
H_{\rm fric} =
  \begin{cases}
  \bar H & \text{(For rigid-FLRW runs)},\\
  H & \text{(inhomogeneous Hubble)}.
  \end{cases}
\end{align}
Now, if one adopts $H_\fric=\bar{H}$ and omits the $\nabla\psi\cdot\nabla\phi$, it reduces to the rigid-FLRW equations.
The results shown in this work are based on the full inhomogeneous Hubble and on keeping the $\nabla\psi\cdot\nabla\phi$ term.

\subsubsection{Local expansion field}
The scalar expansion field $N_{\mathrm{geom}}$ is evolved as 
\begin{equation}
N_{\mathrm{geom}}' = \frac{H}{\bar{H}}.
\label{eq:Ax_prime}
\end{equation}
In a strictly homogeneous FLRW geometry (all inhomogeneity is confined to $\phi$), we have $H = \bar{H}$; in that case, this equation only serves as a consistency check. 
On the other hand, the inhomogeneous case introduces a local expansion history that can vary from cell to cell. 
It also directly provides the 3D grid required to find $\delta{N}$, which we evolve until all grids reach the equal-density hypersurface (or the $\phi = \text{const.}$ surface for single-field inflation in the separate Universe approach).

\subsubsection{Background Raychaudhuri equation}

To evolve the background Hubble rate, we average over the lattice and use the Raychaudhuri equation written in terms of $\mathbb{N}$:
\begin{equation}
\bar H' = -\frac{1}{\bar H}\left[
\frac{1}{2}\,\left\langle \pi_\phi^2 \right\rangle + \frac{1}{6}\,\left\langle e^{-2N_{\rm geom}}|\nabla\phi|^2 \right\rangle + \left\langle C_H \right\rangle
\right],
\label{eq:raychaudhuri_full}
\end{equation}
where the brackets denote a spatial average defined below. 
In the ``separate universe" limit, we typically drop the gradient and curvature contributions and simply keep the kinetic part, which reproduces the usual homogeneous FLRW evolution for $\phi(\mathbb{N})$ and $\bar{H}(\mathbb{N})$.

Time evolution is driven by the background $e$-fold variable $\mathbb{N}$. 
The background equations are solved using a classic fixed-step fourth-order Runge-Kutta (RK4) integrator. 
For the $\delta N$ evaluations, we have implemented both a fixed-step RK4 and an embedded RK4(3) stepper~\cite{Press:2007ipz} with an adaptive step-size controller for stringent error control. 
All dynamical fields $(\phi, \pi_\phi, N_{\mathrm{geom}}, \bar{H})$ are updated consistently at each stage of the integration to maintain the local constraints throughout the simulation.

\subsubsection{Spatial averaging and background quantities}
We define two types of spatial average for a lattice scalar $X_{i,j,k}$:
\begin{enumerate}
  \item Coordinate average
   \begin{align}
   \langle X \rangle_{\rm coord}
   = \frac{1}{\Ng^3} \sum_{i,j,k} X_{ijk}.
   \end{align}

  \item Proper-volume average
   \begin{align}
   \langle X \rangle_{V}
   = \frac{\sum_{i,j,k} e^{3(N_{\mathrm{geom}} - \mathbb{N})} X_{ijk}}
   {\sum_{i,j,k} e^{3(N_{\mathrm{geom}} - \mathbb{N})}}.
   \end{align}
\end{enumerate}
Most background quantities in this work~---~such as $\bar{\phi}$, $\bar{H}$, and the mean energy density $\bar{\rho}$ --- are computed using one of these prescriptions, depending on the physical scenario we want to mimic (e.g., a coordinate average vs a volume-weighted average over expanding patches). 
The Raychaudhuri equation, Eq.~\eqref{eq:raychaudhuri_full}, is written in terms of the chosen volume measure. 
For example, in the $\delta{N}$-runs where we want a `\textit{rigid FLRW}' background consistent with the separate-universe limit, we use coordinate averages and drop the gradient and curvature pieces in Eq.~\eqref{eq:raychaudhuri_full}.
In more general inhomogeneous runs, we retain the gradient and/or curvature terms and may use proper-volume weighting.

\subsection{\label{app:ICs}Initial conditions}
We specify initial data on a spatially flat slice at a chosen \efolds time $\mathbb{N}_{\mathrm{init}}$ (in practice $\mathbb{N}_{\mathrm{init}}=0$ so that $a_{\mathrm{init}}=1$).
The initial configuration for the fundamental evolved variables on the lattice~---~the scalar field $\phi$, its conjugate momenta derivative $\pi_\phi$, and the local logarithmic scale factor $N_{\mathrm{geom}}(\mathbf{x})$ is set as follows:

\subsubsection{Homogeneous background}
We begin from a spatially uniform inflaton configuration,
\begin{equation}
  \phi_{[i,j,k]} = \phi_0, \qquad
  \pi_{\phi,[i,j,k]} = \pi_{\phi,0},
\end{equation}
such that the (initial) background Hubble rate satisfies the Friedmann equation
\begin{equation}
  \bar H_0^2
  \;=\; \frac{1}{3}
  \left( \frac12 \pi_{\phi,0}^2 + V(\phi_0) \right),
\end{equation}
The background \efolds satisfies
$\mathbb{N}(t_{\rm init}) = \mathbb{N}_{\rm init}$, and we set
\begin{equation}
  N_{\mathrm{geom}}(\mathbf{x}) = \mathbb{N}_{\rm init} \equiv 0;\quad(\text{i.e., }\psi(0)=0)
\end{equation}
initially, so that the curvature perturbation in the metric vanishes on the initial slice. 
The precise values of $\phi_0$ and $\pi_{\phi,0}$ are chosen based on the number of \efolds before the end of inflation at which we start our simulation.
When the simulation is initialised in a slow-roll phase, $\pi_{\phi,0}$ can also be obtained from the slow-roll attractor solution.
All inhomogeneity therefore resides in the scalar field sector at the initial time, as we define below.

\subsubsection{\label{app:init_kmodes}Vacuum fluctuations and effective lattice frequency}
To model quantum vacuum fluctuations, we populate the Fourier modes of the field perturbation $\delta\phi$ and its velocity $\delta\pi_{\phi}$ with Gaussian random variables consistent with the adiabatic vacuum in a quasi-de Sitter background.

For our comoving (periodic) box of size $L$ discretised into $\Ng^3$ points, the wavevectors are 
\begin{equation}
  \mathbf{k} = \frac{2\pi}{L}\,\mathbf{n},
  \qquad
  \mathbf{n}=(n_x,n_y,n_z).
\end{equation}
The discrete Laplacian used in the nonlinear evolution does not act with eigenvalue $-k^2$, but with the lattice effective mode $-k_{\rm eff}^2(\mathbf{k})$. 
For the standard second-order nearest-neighbour stencil, this is
\begin{equation}
  k_{\rm eff}^2(\mathbf{k}) =
  \frac{4}{\Delta x^2}
  \left[
    \sin^2\left(\frac{k_x\Delta x}{2}\right)
    + \sin^2\left(\frac{k_y\Delta x}{2}\right)
    + \sin^2\left(\frac{k_z\Delta x}{2}\right)
  \right],
  \label{eq:keff_def}
\end{equation}
where $k_i = 2\pi n_i/L$.
The explicit form of $k_{\mathrm{eff}}(\mathbf{k})$ depends on the stencil chosen for the discrete Laplacian. 
Equation~\eqref{eq:keff_def} applies to the standard second-order nearest-neighbour Laplacian. 
If a higher-order or any other modified stencil is used, $k_{\mathrm{eff}}^2$ must be replaced by the corresponding lattice symbol of that discrete operator.
For modes well below the Nyquist frequency, one has $k_{\rm eff}\simeq |\mathbf{k}|$, whereas near the ultraviolet cutoff, the lattice dispersion deviates from the continuum one. 
In particular, at the corner of the Brillouin zone, the usual continuum lattice momentum is
\begin{equation}
  k_{\rm lat,max}
  = \sqrt{3}\,\frac{\pi}{\Delta x}
  = \frac{\sqrt{3}}{2}\,\Ng\,k_{\min},
  \qquad
  k_{\min}=\frac{2\pi}{L},
\end{equation}
while the discrete Laplacian gives
\begin{equation}
  k_{\rm eff,max}
  = \frac{2\sqrt{3}}{\Delta x}
  = \frac{2}{\pi}\,k_{\rm lat,max}.
\end{equation}
In all cases, we use the same lattice dispersion relation in the vacuum initialisation and in the subsequent nonlinear evolution, so that the ultraviolet behaviour of the initial spectrum is consistent with the dynamics implemented on the grid~\cite{Stamatopoulos:2012np,Caravano:2021pgc,Caravano:2022epk}.

We define the effective mass and mode frequency at $\mathbb{N}_{\rm init}$ by
\begin{equation}
  m_{\rm eff}^2 \equiv V_{\phi\phi}(\phi_0),
  \qquad
  \omega_k^2 \equiv k_{\rm eff}^2 + a^2 m_{\rm eff}^2,
\end{equation}
where $V_{\phi\phi}$ is evaluated on the homogeneous background. 
With our initialisation $a_0=1$, the comoving and physical momenta coincide, and the above dispersion relation is directly applicable.

For each independent lattice mode, we draw a complex Gaussian variable $\mathcal{G}_\phi(\mathbf{k})$ with $\langle \mathcal{G}_\phi \rangle = 0$ and $\langle |\mathcal{G}_\phi|^2 \rangle = 1$, imposing the usual reality condition $\delta\phi_{-\mathbf{k}} = \delta\phi_{\mathbf{k}}^\ast$.
We then set
\begin{equation}
  \delta\phi_{\mathbf{k}}(\mathbb{N}_{\rm init})
  \;=\; W(k_{\rm eff}) \,
       \sqrt{\frac{1}{2\,\omega_k\,V}} \;
       \mathcal{G}_\phi(\mathbf{k}),
  \label{eq:IC-dphi}
\end{equation}
where $V = L^3$ is the comoving volume and $W(k_{\rm eff})$ is a smooth UV window that, if applied, suppresses modes near the Nyquist frequency to control aliasing.

The initial field velocity is chosen to satisfy the adiabatic condition for a massive mode in a slowly varying quasi-de Sitter background. 
Denoting the background Hubble rate at $\mathbb{N}_{\rm init}$ by $H_0$, we take
\begin{equation}
  \delta\pi_{\phi,{\mathbf{k}}}(\mathbb{N}_{\rm init})
  \;=\; \bigl(-H_0 - i\,\omega_k\bigr)\,
        \delta\phi_{\mathbf{k}}(\mathbb{N}_{\rm init}),
  \label{eq:IC-dphidot}
\end{equation}
for generic complex modes. This choice reproduces the flat-space vacuum correlators,
\begin{equation}
  \langle |\delta\phi_{\mathbf{k}}|^2 \rangle
  = \frac{W(k_{\rm eff})^2}{2\,\omega_k\,V},
  \qquad
  \langle |\delta\pi_{\phi,{\mathbf{k}}}|^2 \rangle
  \simeq \frac{\omega_k\,W(k_{\rm eff})^2}{2\,V},
\end{equation}
and ensures the correct phase relation between $\delta\phi_{\mathbf{k}}$ and $\delta\pi_{\phi,{\mathbf{k}}}$ for a positive-frequency (Bunch--Davies) mode.
The $-H_0\,\delta\phi_{\mathbf{k}}$ term in Eq.~\eqref{eq:IC-dphidot} approximates the effect of Hubble friction on subhorizon oscillations.

On the special self-conjugate modes of the real-to-complex \ac{FFT} half-space (e.g., $k_z=0$ and $k_z = k_{\rm Nyquist}$ planes), we draw only a real Gaussian amplitude, enforce $\delta\phi_{\mathbf{k}}$ to be real, and use the corresponding real part of Eq.~\eqref{eq:IC-dphidot}. 
In all cases, the Hermitian symmetry in Fourier space is enforced exactly so that the real-space fields are strictly real.

In addition to this `WKB' initialiser, we have implemented an alternative scheme based on the exact Hankel-function solution of the linear mode equation in a quasi-de Sitter background. 
In that scheme, the mode function and its time derivative for each $\mathbf{k}$ are obtained from the analytic slow-roll Hankel solution evaluated at $\mathbb{N}_{\rm init}$, and the random Gaussian amplitudes simply set the overall stochastic normalisation. 
The results that we showed in this work are using the WKB initialisation \eqref{eq:IC-dphi}–\eqref{eq:IC-dphidot}.

After populating all the Fourier modes, we ensure that the zero mode is explicitly vanished,
\begin{equation}
  \delta\phi_{\mathbf{k}=\mathbf{0}} =
  \delta\pi_{\phi,{\mathbf{k}=\mathbf{0}}} = 0,
\end{equation}
ensuring that the homogeneous background $(\phi_0,\pi_{\phi,0})$ is not double-counted.
Finally, we perform an inverse Fourier transform to obtain the real-space fluctuations and initialize the fields at each lattice point at $N_{\mathrm{init}}=0$:
\begin{equation}
  \phi(\mathbf{x})   = \phi_0   + \delta\phi(\mathbf{x}), \qquad
  \pi_{\phi}(\mathbf{x}) = \pi_{\phi,0} + \delta\pi_{\phi}(\mathbf{x}).
\end{equation}

\subsubsection{\label{sec:latt_con}Lattice consistency and constraints}

The initialisation procedure is constructed to preserve both the periodic boundary conditions of the lattice and the reality of the fields. As an additional validation, we monitor the consistency between the background expansion and the local generalised Friedmann constraint.
In the numerical implementation, the background Hubble rate, $\bar H(\mathbb{N})$, is evolved using the volume-averaged Raychaudhuri equation. In parallel, the local Hubble rate, $H(\mathbf{x}, \mathbb{N})$, is reconstructed pointwise from the matter energy density, supplemented by the curvature correction term $C_H$, as defined in Eq.~\eqref{eq:Hx2_def}. The consistency between these two constructions is ensured by the volume-averaged relation
\begin{equation}
\bar H = \langle H \rangle_V,
\label{eq:const}
\end{equation}
provided that the background scale factor is defined in terms of the physical volume of the simulation domain, as detailed in Sec.~\ref{app:volavg}.
In this sense, the simulation tracks not only the local field dynamics but also the compatibility between the averaged Raychaudhuri evolution and the local Hamiltonian constraint. 
In the rigid-FLRW limit, this reduces to the familiar background energy-consistency check, while in the inhomogeneous case, it provides a more general consistency check between the averaged background evolution and the local generalised Friedmann constraint. 
In all our runs, we found that this consistency relation is satisfied to floating-point precision throughout the simulation.

We do not explicitly solve the $0i$ Einstein (momentum) constraint at $\mathbb{N}_{\mathrm{init}}$; instead, the resulting initial residual of the momentum constraint is monitored during the evolution. In practice, this residual decays rapidly as the system relaxes to the self-consistent, inhomogeneous, slow-roll solution and remains small throughout the physically relevant part of the simulation.

\subsubsection{Momentum constraint monitoring}
To monitor the Einstein 0i (momentum) constraint, we define the residual vector field
\begin{equation}
\mathfrak{R}_i \equiv \partial_i H + \frac{1}{2\Mpl^2}\,\dot\phi\,\partial_i\phi,
\label{eq:momentum_constraint}
\end{equation}
We also define,
\begin{align}
L_i \equiv \partial_i H, \qquad
R_i \equiv -\frac{1}{2M_{\rm Pl}^2}\,\dot\phi\,\partial_i\phi,
\end{align}

On the lattice, we compute the discrete gradient of $H$ and $\phi$ with the same central finite differences as in the equations of motion, and construct the proper-volume-weighted \ac{rms} $\langle |\mathfrak{R}|^2\rangle_V^{1/2}$,
and a dimensionless normalised measure
  \begin{align}\label{eq:normR}
   \text{norm}(\mathfrak R) \equiv \frac{\langle|\mathfrak R|^2\rangle_V^{1/2}}{\langle|L_i|^2\rangle_V^{1/2} + \langle|R_i|^2\rangle_V^{1/2}}.
  \end{align}

\bibliographystyle{JHEP}
\bibliography{biblio}

\end{document}